\documentclass{iopart}
\usepackage{amssymb}
\usepackage{graphicx}
\usepackage{float}
\usepackage{geometry}

\makeatletter
\@namedef{ver@amsmath.sty}{}
\makeatother
\usepackage{amsmath}

\begin{document}
\title{Clogging, Diode and Collective Effects of Skyrmions in Funnel Geometries}
\author{J. C. Bellizotti Souza$^1$, 
            N. P. Vizarim$^2$, 
            C. J. O. Reichhardt$^3$, 
            C. Reichhardt$^3$
            and P. A. Venegas$^1$}
\address{$^1$ Departamento de Física, Faculdade de Ciências, Unesp-Universidade Estadual Paulista, CP 473, 17033-360 Bauru, SP, Brazil}
\address{$^2$ POSMAT - Programa de Pós-Graduação em Ciência e Tecnologia de Materiais, Faculdade de Ciências, Universidade Estadual Paulista - UNESP, Bauru, SP, CP 473, 17033-360, Brazil}
\address{$^3$ Theoretical Division and Center for Nonlinear Studies, Los Alamos National Laboratory, Los Alamos, New Mexico 87545, USA
}
\ead{cjrx@lanl.gov, nicolas.vizarim@unesp.br}

\begin{abstract}
Using a particle-based model, we examine the collective dynamics of skyrmions interacting with a funnel potential under dc driving as the skyrmion density and relative strength of the Magnus and damping terms are varied. For driving in the easy direction, we find that increasing the skyrmion density reduces the average skyrmion velocity due to jamming of skyrmions near the funnel opening, while the Magnus force causes skyrmions to accumulate on one side of the funnel array. For driving in the hard direction, there is a critical skyrmion density below which the skyrmions become trapped. Above this critical value, a clogging effect appears with multiple depinning and repinning states where the skyrmions can rearrange into different clogged configurations, while at higher drives, the velocity-force curves become continuous. When skyrmions pile up near the funnel opening,
the effective size of the opening is reduced and the passage of other skyrmions is blocked by the repulsive skyrmion-skyrmion interactions. We observe a strong diode effect in which the critical depinning force is higher and the velocity response is smaller for hard direction driving. As the ratio of Magnus force to dissipative term is varied, the skyrmion velocity varies in a non-linear and non-monotonic way due to the pile up of skyrmions on one side of the funnels. At high Magnus forces, the clogging effect for hard direction driving is diminished.
\end{abstract}

\maketitle
\vskip 2pc

\section{Introduction}
    
The dynamical behavior of vortices in superconductors, Wigner crystals,
and colloidal particles subjected to confined 
geometries such as one-dimensional (1D) channels or constrictions 
\cite{koppl_layer_2006,dobrovolskiy_electrical_2012,barrozo_model_2009}, 
bottlenecks \cite{piacente_pinning_2005,yu_vortex_2010,rees_commensurability-dependent_2012,zimmermann_flow_2016}, 
or asymmetric funnel walls \cite{wambaugh_superconducting_1999,yu_asymmetric_2007, reichhardt_commensurability_2010,vlasko-vlasov_jamming_2013,karapetrov_evidence_2012}
has been extensively investigated in the past years.  
In such systems, it is possible for a portion of the
particles to become
pinned by the confining walls, 
thereby 
hindering the motion of other particles due to
the particle-particle interactions.
Clogging is a generic phenomenon
that has been studied for particles with contact interactions
such as granular matter
both with and without friction flowing
through an aperture or orifice, where
when the density is high enough,
the flow can stop
\cite{redner_clogging_2000,zuriguel_clogging_2014,to_jamming_2001,helbing_simulating_2000,nguyen_clogging_2017}.
The clogging is probabilistic,
and 
the system can flow for some time
until the particles reach a certain configuration that stops the flow. 
   
For systems of particles interacting with funnel arrays, there  
have been a variety of studies
describing
diode, ratchet, clogging and jamming effects for superconducting vortices 
 \cite{reichhardt_commensurability_2010,reichhardt_jamming_2010,olson_reichhardt_vortex_2013,vlasko-vlasov_jamming_2013,wambaugh_superconducting_1999,villegas_experimental_2005,villegas_superconducting_2003},
colloidal particles \cite{reichhardt_clogging_2018}, and active matter
\cite{martinez_trapping_2020,wan_rectification_2008,berdakin_influence_2013}.
In the case of superconducting vortices and colloids,
the interactions are longer range compared to the contact forces
found in granular matter.
A periodic array of funnels
can also produce
commensurability effects
in which the depinning threshold
is significantly increased when the number of particles is an integer
multiple of the
number of plaquettes in the system
\cite{olson_reichhardt_vortex_2013,reichhardt_commensurability_2010,reichhardt_clogging_2018}.
Due to the asymmetry of the funnel geometry,
the depinning and dynamics
depend strongly
on whether the drive is applied along the easy or hard direction of
motion through the funnel.
For driving in the easy direction, the velocity-force curves are generally
smooth, while for driving
in the hard direction,
clogging dynamics can occur when particles  accumulate along the
funnel edges and block the flow.

Recently, a particle-based model of magnetic skyrmions was 
investigated in the presence of funnel
geometries \cite{Migita20,souza_skyrmion_2021}.
Here
a single skyrmion was subjected to
ac driving,
and it was possible to achieve precise control
of the skyrmion motion
through a ratchet effect
\cite{souza_skyrmion_2021}.
The ratcheting can be induced with ac driving in the particle-based
model, but could be produced by
an oscillating field that changes the size of the skyrmion
in a continuum model \cite{Migita20}.
In the present work we 
examine collective behaviors including diode and jamming  effects
for assemblies
of skyrmions interacting with a linear array 
of asymmetric funnels.

Skyrmions are spin textures wrapping a sphere pointing in all directions to form a 
topologically
stable particle-like object \cite{muhlbauer_skyrmion_2009,fert_magnetic_2017,fert_skyrmions_2013}
that can be set into motion by the application of a
spin polarized current 
\cite{jonietz_spin_2010,schulz_emergent_2012,yu_skyrmion_2012}.
In the presence of external drives, skyrmions can show a depinning threshold
similar to 
that found
for superconducting vortices. The key difference between skyrmions and other overdamped 
particles is the presence
of the Magnus force \cite{nagaosa_topological_2013,iwasaki_universal_2013}.
This non-dissipative force produces a skyrmion velocity contribution 
that is perpendicular to the external applied current.
In the case of a clean sample without defects, the skyrmions move
at an angle with respect to the transport force
known as the skyrmion Hall angle,
$\theta_{sk}^{\rm int}$
\cite{nagaosa_topological_2013,fert_magnetic_2017,jiang_direct_2017}.
Experimentally, the skyrmion Hall angle ranges
from a few degrees up to values very close to
$90^{\circ}$ depending on the system parameters \cite{jiang_direct_2017,litzius_skyrmion_2017}.
    
Recent advances in creating nanoengineered materials, 
where defects of different sizes and shapes can be embedded in a sample,
permit the creation of 
complex structures that can interact
with and guide skyrmions inside the material.
In this work we 
use a particle-based simulation to describe the dynamical behavior of multiple skyrmions 
interacting with a linear array of asymmetric funnels under the influence of a dc drive. 
In previous works investigating single skyrmion dynamics under the influence of 
periodic and asymmetric potentials,
we showed that the Magnus term can significantly influence
the dynamics and can produce quantized directions of motion 
\cite{reichhardt_quantized_2015,vizarim_directional_2021,feilhauer_controlled_2020}, novel 
ratchet effects \cite{gobel_skyrmion_2021,vizarim_skyrmion_2020,souza_skyrmion_2021,reichhardt_magnus-induced_2015,ma_reversible_2017} and possibilities for controlled skyrmion 
motion \cite{vizarim_guided_2021}.
When multiple interacting skyrmions are present in a 
sample, the skyrmion-skyrmion interactions can compete with the effects
of the Magnus force.
It was shown that in the presence of random disorder,
an assembly of skyrmions flows in riverlike channels 
with individual skyrmions switching between pinned and moving states 
\cite{reichhardt_noise_2016,montoya_spin-orbit_2018},
while for multiple skyrmions in 
periodic pinning,
moving segregated states appear with
clustering that arises 
due to the velocity dependence of the skyrmion
Hall angle in the presence of strong pinning 
\cite{reichhardt_nonequilibrium_2018}.

In this work, we investigate the collective interactions
of skyrmions in a linear array of asymmetric funnels where a
dc transport force is applied along the easy or hard direction.
At low skyrmion density, for driving along the easy axis
the dynamics is similar to that found
in the single skyrmion case, where the average velocity increases 
linearly with increasing drive.
As the skyrmion density increases,
collective effects emerge and 
the skyrmions show a  jamming effect  near the funnel opening
associated with
a reduction in the average skyrmion velocity.
For drives applied along the hard direction,
below a critical skyrmion density
there is no net skyrmion motion,
while at higher densities the
skyrmions can exhibit
more than one finite depinning threshold
due to reentrant pinning phases (RPPs)
produced
by clogging effects. In the RPPs, the skyrmion motion is 
interrupted due to a pile up of skyrmions inside funnel plaquettes, leading to
formation of a clogged 
state
where the skyrmions cannot move until
a second depinning threshold
is reached that opens
the flow of skyrmions through the funnels once again.
The clogging effect is probabilistic in nature,
as indicated by the fact that the
depinning and reentrant pinning transitions
fall at different driving forces for
different realizations.
At higher drives,
the velocity-force
curves become continuous,
and we map the transition from clogging 
behavior to continuous flow as
a function of skyrmion density and driving force.
A skyrmion diode effect is produced by the different responses of
the driven skyrmions under different driving directions.
We have also studied the effect of varying the ratio of the
Magnus term to the dissipative term for both directions of driving,
and find both non-linear and non-monotonic behavior.
For hard direction driving,
increasing the Magnus term can diminish and
destroy the RPPs
by reducing the tendency of the skyrmions to clog.

    \section{Simulation}
\begin{figure}[ht]
  \begin{center}
    \includegraphics[width=0.5\columnwidth]{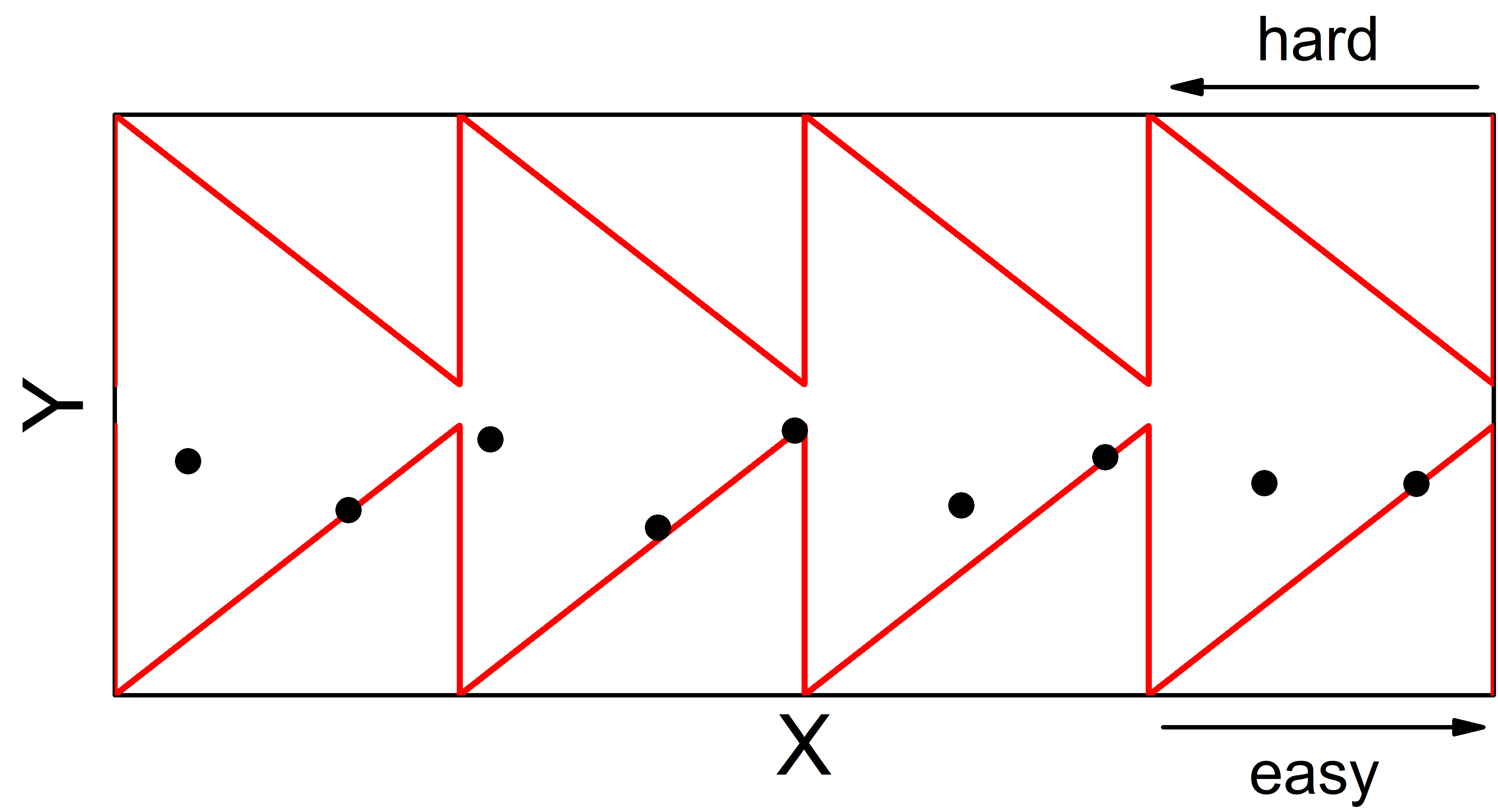}
  \end{center}
\caption{Illustration of the linear array of funnel potentials used in this work. Red lines are the repulsive barrier walls of the funnel geometry
  and black dots are skyrmions that are subjected to both damping and Magnus terms as well as a dc drive applied along the $+x$ or $-x$ direction.
  The hard and easy directions of motion are labeled with arrows.
}
\label{Fig1}
\end{figure}

We simulate the collective effects of skyrmions under the influence of a linear array of funnel potentials aligned along the $x$ direction, as shown in Fig. \ref{Fig1}.
The simulation box has dimensions $L_x\times L_y$
with periodic boundary conditions in the
    $x$ direction.
    
The skyrmions interact with each other, with the
repulsive barrier walls that form the funnel geometry,
and with an applied dc drive. 
The skyrmion dynamics is governed by equation (\ref{vels}) \cite{lin_particle_2013}:
    
    \begin{equation}\label{vels}
        \alpha_d\mathbf{v}_{i}+\alpha_m\hat{z}\times\mathbf{v}_{i}=\mathbf{F}_{i}^{ss}+\mathbf{F}_{W}+\mathbf{F}^{D}
    \end{equation}

Here, $\mathbf{v}_{i}$ is the velocity of the $i$th skyrmion. The first term on the left 
is the damping 
term that arises 
from the spin precession
and dissipation of electrons in the skyrmion core,
where $\alpha_d$ is the damping constant. The second term
is the Magnus term
produced by
gyroscopic effects, 
where $\alpha_m$ is the Magnus constant. 
Throughout this work we use the normalization $\alpha_d^2+\alpha_m^2=1.0$.

On the right side of equation (\ref{vels}), 
the first term is the repulsive skyrmion-skyrmion force given by the expression
$\mathbf{F}^{ss}=\sum^{N}_{i}K_{1}(r_{ij}/\xi)\hat{\mathbf{r}}_{ij}$,
where $\xi$ is the screening length which we take to be $1.0$
in dimensionless units,
$r_{ij}=|\mathbf{r}_{i}-\mathbf{r}_{j}|$ 
is the distance between skyrmions $i$ and $j$, and the unit vector 
$\hat{\mathbf{r}}_{ij}=(\mathbf{r}_{i}-\mathbf{r}_{j})/r_{ij}$.
For computational efficiency, we cut off
the interaction beyond $r_{ij}=6.0$ where the
Bessel interaction becomes negligible.
The sample size is fixed to $L_x=35\xi$ and
$L_y=14\xi$.
The second term on the right side is the funnel wall potential,
which is modeled using a 
Gaussian function given by $U_W=U_0e^{-r_{iw}^{2}/a_0^{2}}$, where
$U_0$ is the potential strength, $r_{iw}$ is the distance between the skyrmion $i$ and the
wall, and $a_0$ is the 
funnel wall thickness.
The repulsive force exerted by the wall is 
    $\mathbf{F}_W=-\nabla U_W$,
giving $\mathbf{F}_W=F_0r_{iw}e^{-r_{iw}^2/a_0^2}\hat{\mathbf{r}}_{iw}$ 
where $F_0=2U_0/a_0^2$. For this work
we fix $a_0=0.02\xi$ and $U_0=1.0$. The number of funnels is fixed at $N_F=4$ and the funnel opening is set to $O=1.0\xi$.
The skyrmion density is given by
$\rho_{sk}=N_{sk}/L_xL_y$, where $N_{sk}$ is the number
of skyrmions in the sample.
This quantity is normalized in terms of $1/\xi^{2}$.

The last term on the right side of equation (\ref{vels})
is the transport dc driving force 
$\mathbf{F}^{D}=F^{D}\hat{\mathbf{d}}$
with $\hat{\mathbf{d}}=\pm\hat{\mathbf{x}}$.
Here 
$+\hat{\mathbf{x}}$
is the easy driving direction
and $-\hat{\mathbf{x}}$ is the hard driving direction.
The external drive is increased in small steps of $\delta F^D=0.001$ and
we spend
$5\times10^5$ simulation time steps
on each step to evaluate the time averaged 
velocity measurement,
$\left\langle V_x\right\rangle=\left\langle\mathbf{v}\cdot\hat{\mathbf{x}}\right\rangle$.

\section{Easy direction driving}
\begin{figure}[ht]
\begin{center}
\includegraphics[width=0.35\columnwidth]{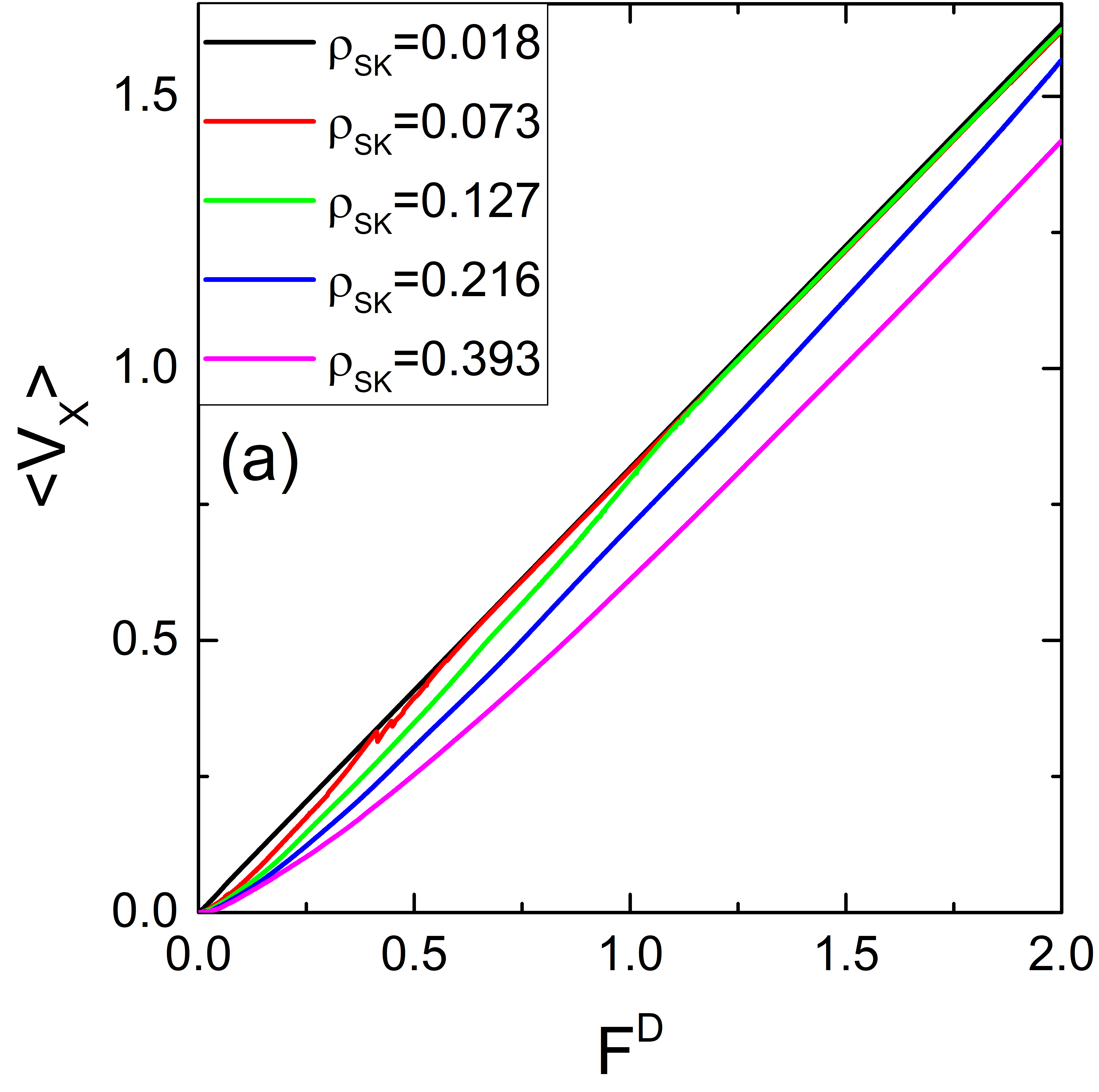}%
\includegraphics[width=0.35\columnwidth]{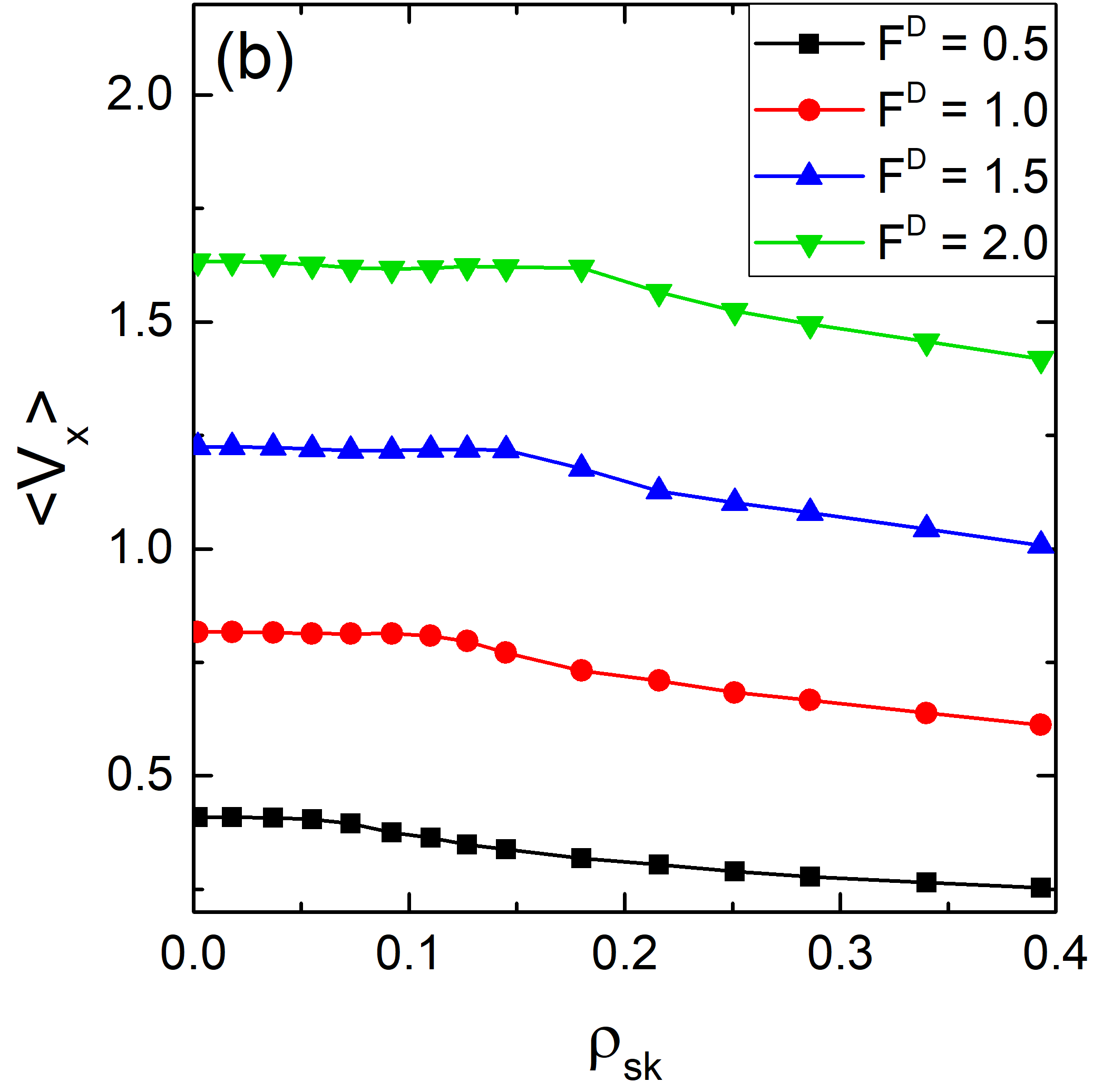}
        \end{center}
\caption{
Results for a sample with
$\alpha_m/\alpha_d = 0.5$ and
easy or $+x$ direction driving.
(a) $\langle V_x\rangle$ vs drive amplitude $F^D$ for
skyrmion densities $\rho_{sk}=0.018$, 0.073, 0.127, 0.216, and 0.393.
For high values of $F^D$, the curves become linear.
$\langle V_x\rangle$ decreases with increasing skyrmion density,
showing a clear 
effect of the collective behavior.
(b) $\langle V_x \rangle$ vs $\rho_{sk}$
for $F^D$=0.5, 1.0, 1.5, and 2.0.
        }
        \label{Fig2}
    \end{figure}

First we consider the case where the external dc drive is
applied along the easy or $+x$ direction.
In Fig.~\ref{Fig2}(a) we plot the average skyrmion velocity 
$\langle V_x \rangle$ as a function of the external drive $F^D$ for several values of the skyrmion
density $\rho_{sk}$
in a sample with $\alpha_m/\alpha_d=0.5$.
Figure~\ref{Fig2}(b) shows $\langle V_x \rangle$
versus $\rho_{sk}$ for selected values of $F^D$.

\begin{figure}[ht]
\begin{center}
  \includegraphics[width=0.6\columnwidth]{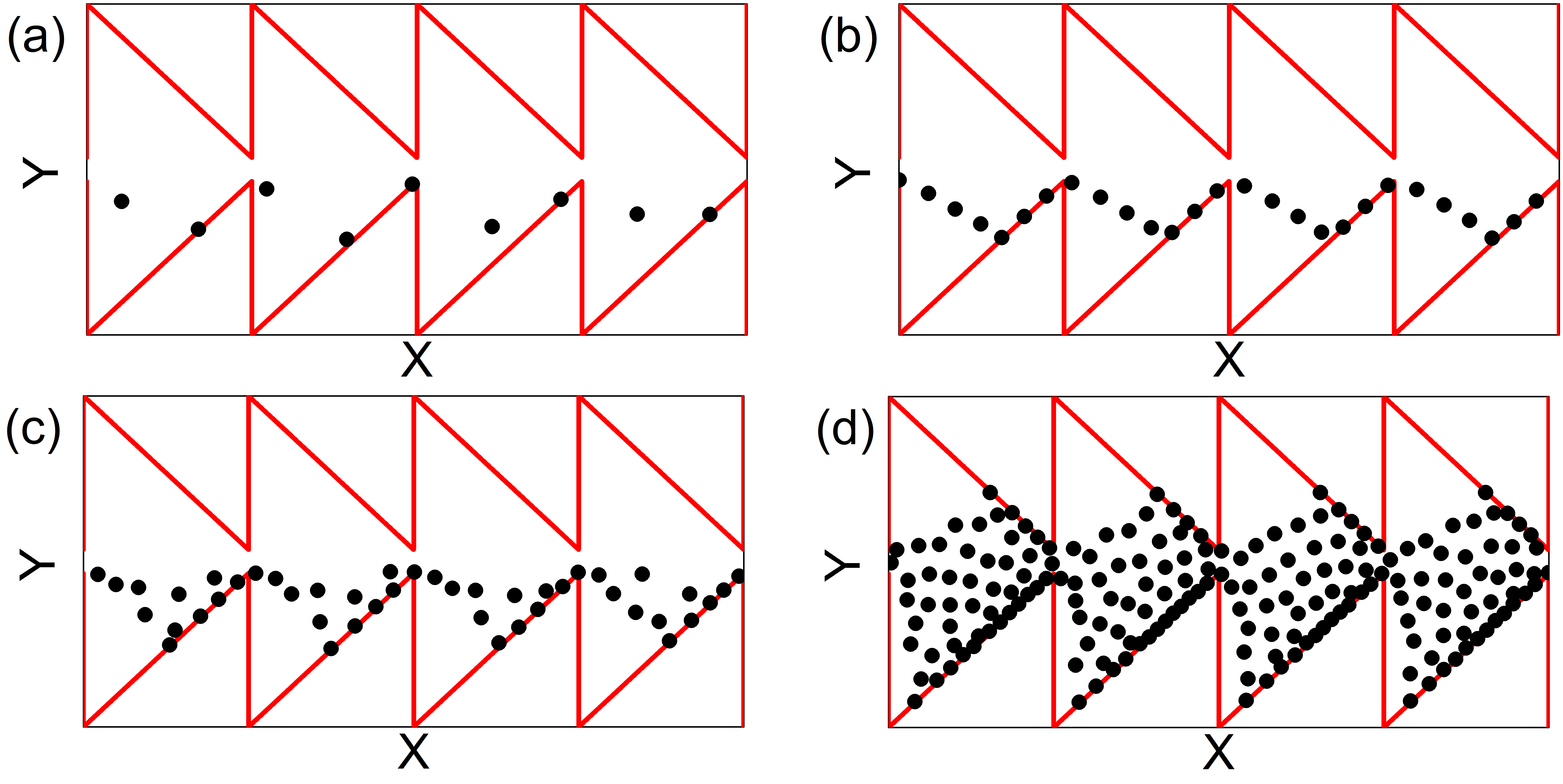}
  \end{center}
\caption{
Snapshots of the skyrmion positions for the system in
Fig.~\ref{Fig2} with $\alpha_m/\alpha_d=0.5$ and $+x$
driving at $F^D=1.5$.
Red lines are the funnel walls
and black dots are skyrmions.
The skyrmion density $\rho_{sk}=$ (a) 0.018, (b) 0.055, (c) 0.092 and (d) 0.393.
As $\rho_{sk}$ increases, there is a decrease in
$\langle V_x\rangle$ due to the competition among skyrmions to pass through
the funnel opening.
There is also a buildup of skyrmion density
in the lower parts of the funnel due to the Magnus effect.       
}
\label{Fig3}
\end{figure}

In general, for low values of $\rho_{sk}$ the velocity curve
$\langle V_x \rangle$ increases linearly with increasing external drive.
The flowing skyrmions are widely spaced,
as shown in Fig.~\ref{Fig3}(a),
making collective effects almost negligible.
The response is similar to that found for a
single skyrmion,
where a linear dependence of $\langle V_x\rangle$ on $F^D$ 
is 
expected.
As more skyrmions are added to the sample, the skyrmion-skyrmion interactions 
become 
important and the skyrmions begin to compete with each other to pass
through the funnel opening.
This competition
results in a nonlinear $\langle V_x \rangle$ curve,
as shown in Fig.~\ref{Fig2}(a). 
For each value of $F^D$ there is a critical value of $\rho_{sk}$
above which
collective effects appear and
the average velocity $\langle V_x \rangle$ begins to drop,
as shown in Fig.~\ref{Fig2}(b).
With increasing $\rho_{sk}$,
skyrmions accumulate near the funnel opening,
leading to a jamming of the skyrmions
and a corresponding reduction in
the average velocity.
    
In Fig.~\ref{Fig3} we show some snapshots of the moving skyrmions
for the system in Fig.~\ref{Fig2}.
At $\rho_{sk}=0.018$ in Fig.~\ref{Fig3}(a),
the skyrmions interact only weakly and
the skyrmion flow is very similar to 
that found in the single skyrmion limit.
For $\rho_{sk}=0.055$, illustrated in Fig.~\ref{Fig3}(b),
the skyrmions flow in a closely spaced line, 
increasing the relevance of collective effects.
In Fig.~\ref{Fig3}(c) at $\rho_{sk}=0.092$, the skyrmions can 
no longer remain stable in a single line of flow,
and competition between skyrmions to pass through the funnel
openings begins to appear.
For
$\rho_{sk}=0.393$
in Fig.~\ref{Fig3}(d), multiple skyrmions agglomerate 
near the funnel opening, decreasing its effective width through their
repulsive interactions with approaching skyrmions.
This is associated with
a significant reduction in the average velocity and can lead to
the formation of a
\textit{jammed state}
of skyrmions near the funnel opening.
Here the skyrmions flow almost like a fluid
trying to
pass through a tight orifice.
Due to the  Magnus term, when the drive is applied along the $+x$ 
direction, the skyrmions move at an angle with respect to the drive given
by the skyrmion Hall angle,
$\theta_{sk}^{\rm int}=\arctan(\alpha_m/\alpha_d)=-26.56^{\circ}$,
producing
a higher concentration of skyrmions in the lower part of the funnel array,
as can be seen clearly in Fig.~\ref{Fig3}(c,d).

\section{Hard direction driving}

\begin{figure}[ht]
\begin{center}
     \includegraphics[width=0.35\columnwidth]{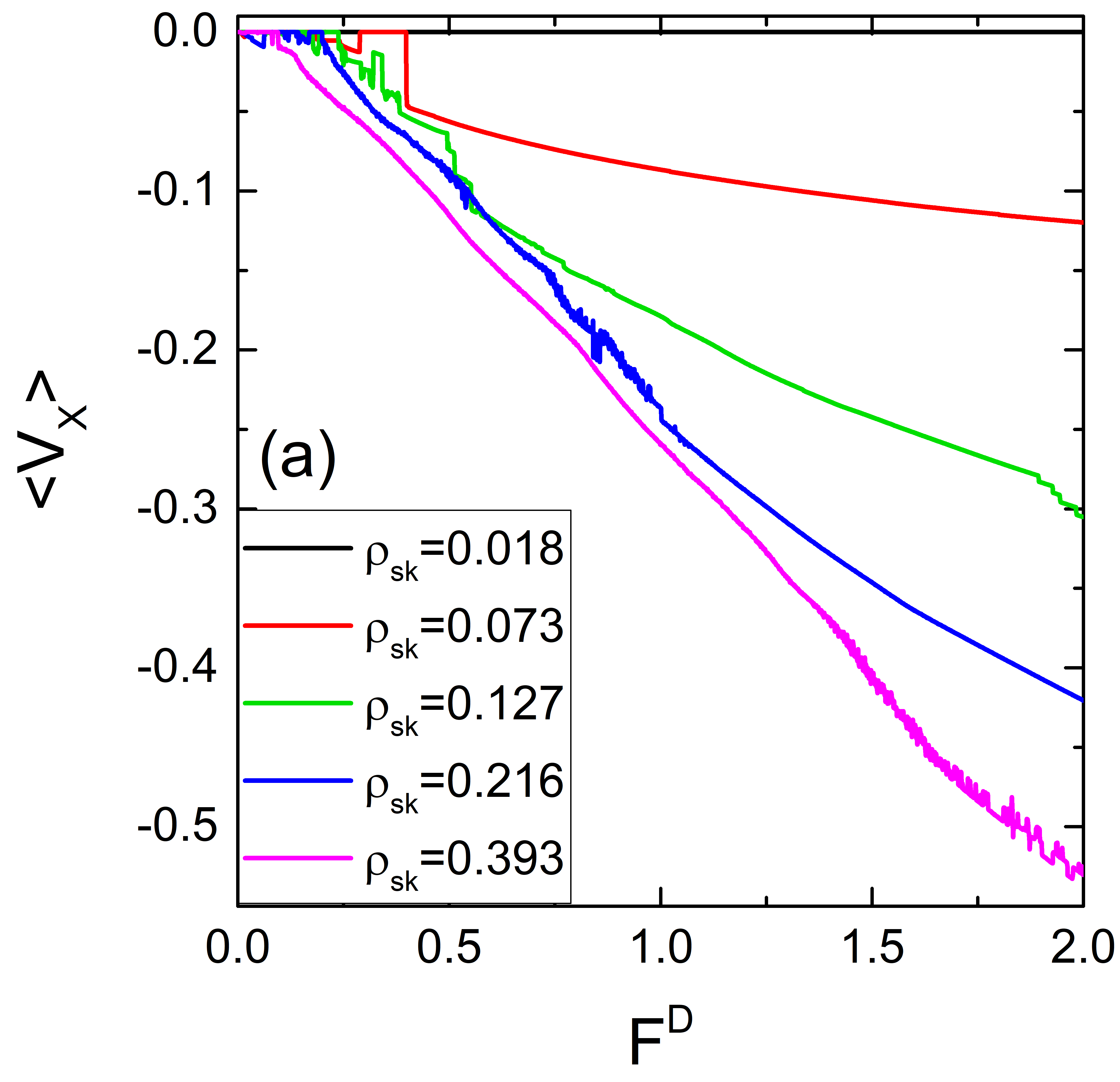}%
     \includegraphics[width=0.35\columnwidth]{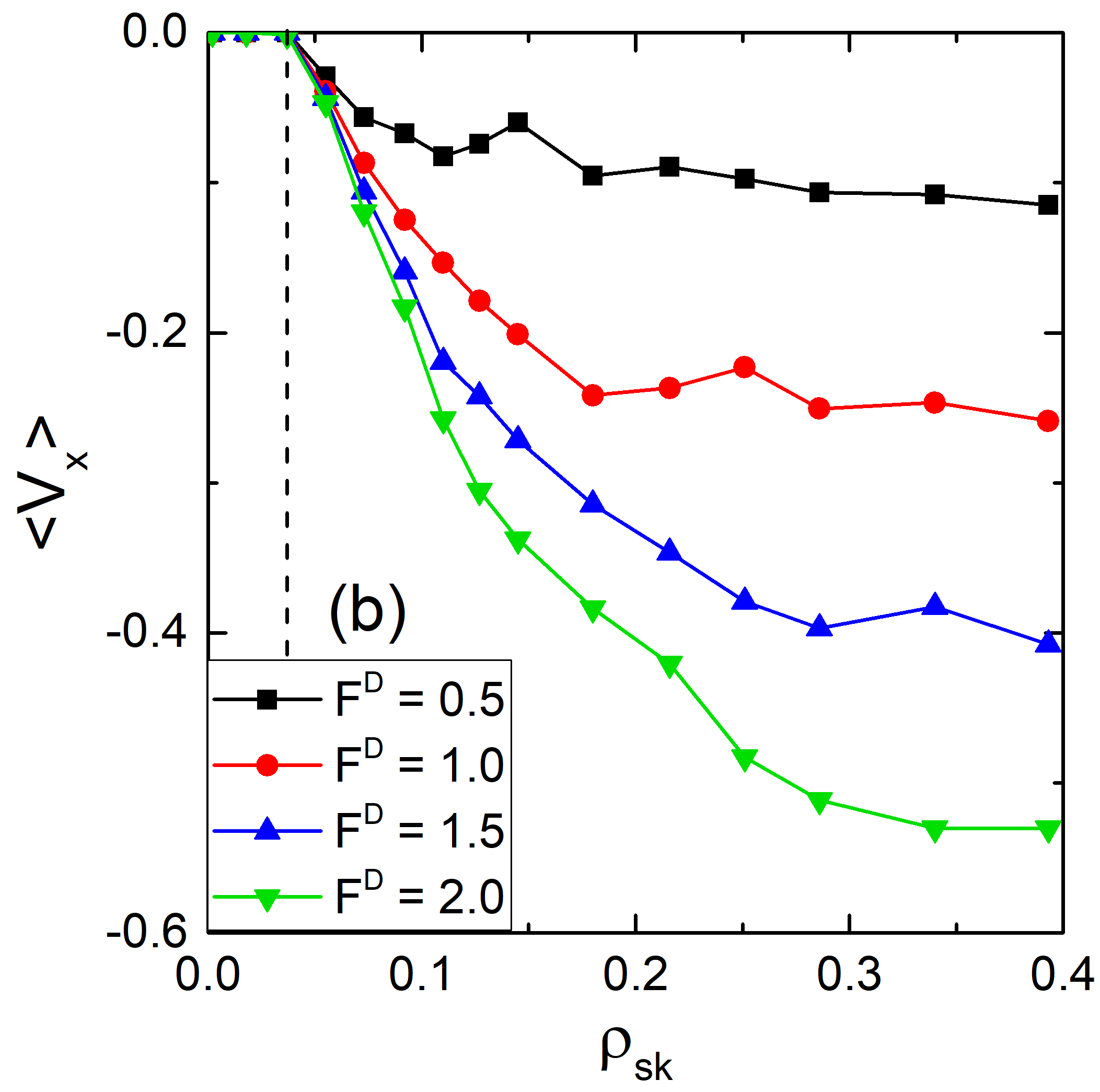}
\end{center}
\caption{
Results for a sample with $\alpha_m/\alpha_d = 0.5$ and
hard or $-x$ direction driving.
(a) $\langle V_x\rangle$ vs $F^D$ for
$\rho_{sk}$=0.018, 0.073, 0.127, 0.216, and 0.393. 
Some of the curves show non-monotonic behavior.
(b) $\langle V_x \rangle$ vs $\rho_{sk}$ for
$F^D=0.5$, 1.0, 1.5, and 2.0,
where we find a critical density value (vertical dashed line) of
$\rho_{sk}^{\rm crit} = 0.037$
}
\label{Fig4}
\end{figure}

We next apply the drive along the hard
or $-x$ direction for a sample with $\alpha_m/\alpha_d=0.5$.
In Fig.~\ref{Fig4}(a) we show
$\langle V_x\rangle$ versus $F^D$ at different skyrmion densities,
and in Fig.~\ref{Fig4}(b) we plot
$\langle V_x \rangle$ versus $\rho_{sk}$ for selected values of $F^D$. 

\begin{figure}[ht]
\begin{center}
  \includegraphics[width=0.6\columnwidth]{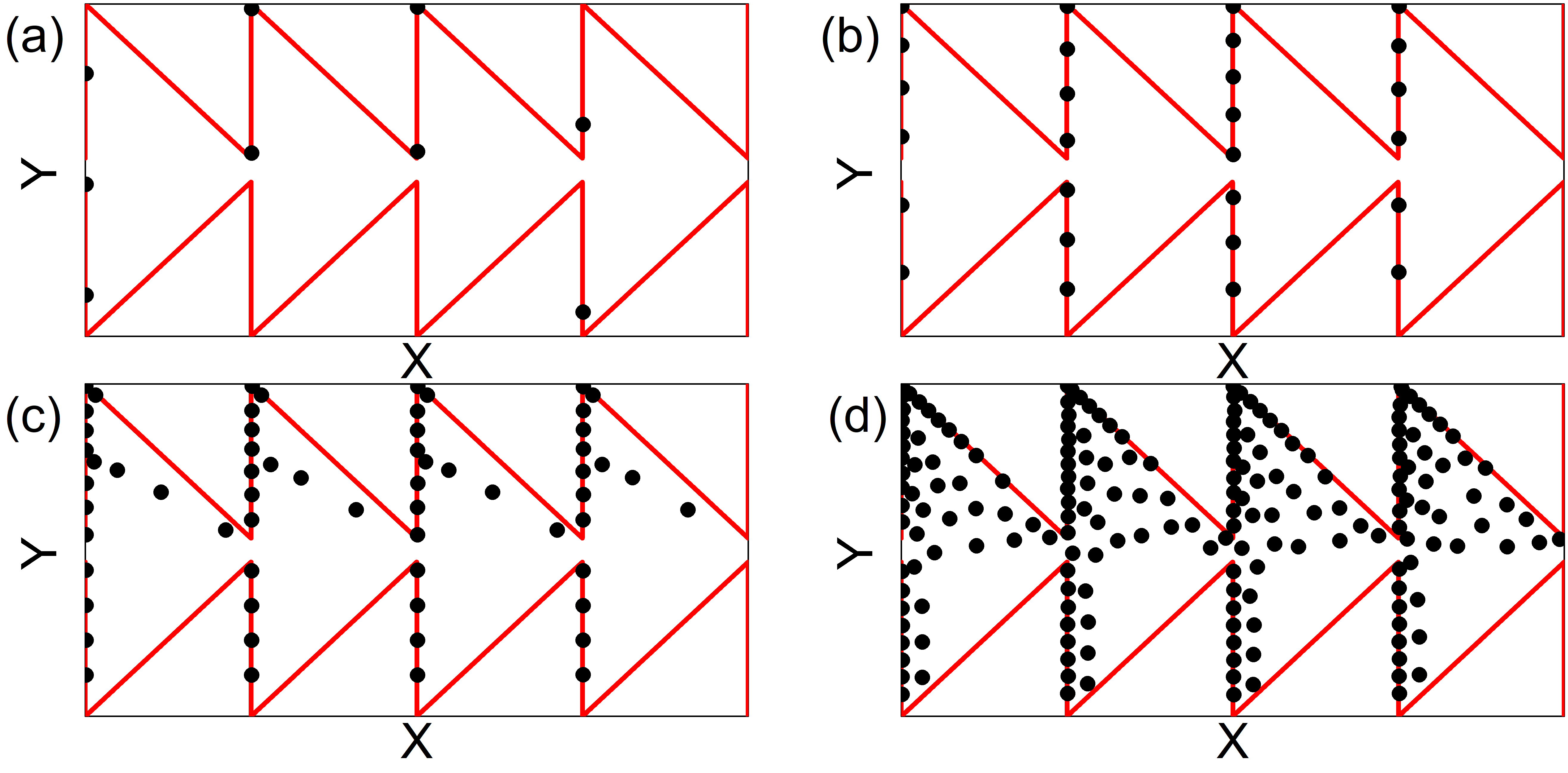}
\end{center}  
\caption{
Snapshots of the skyrmion positions for the system in
Fig.~\ref{Fig4} with $\alpha_m/\alpha_d=0.5$ and $-x$ driving
at $F^D=1.5$. 
Red lines are the funnel walls
and black dots are skyrmions.
The skyrmion density $\rho_{sk}=$ (a) 0.018, (b) 0.055, (c) 0.127 and (d) 0.393.
}
\label{Fig7}
\end{figure}

\begin{figure}[ht]
\begin{center}
  \includegraphics[width=0.35\columnwidth]{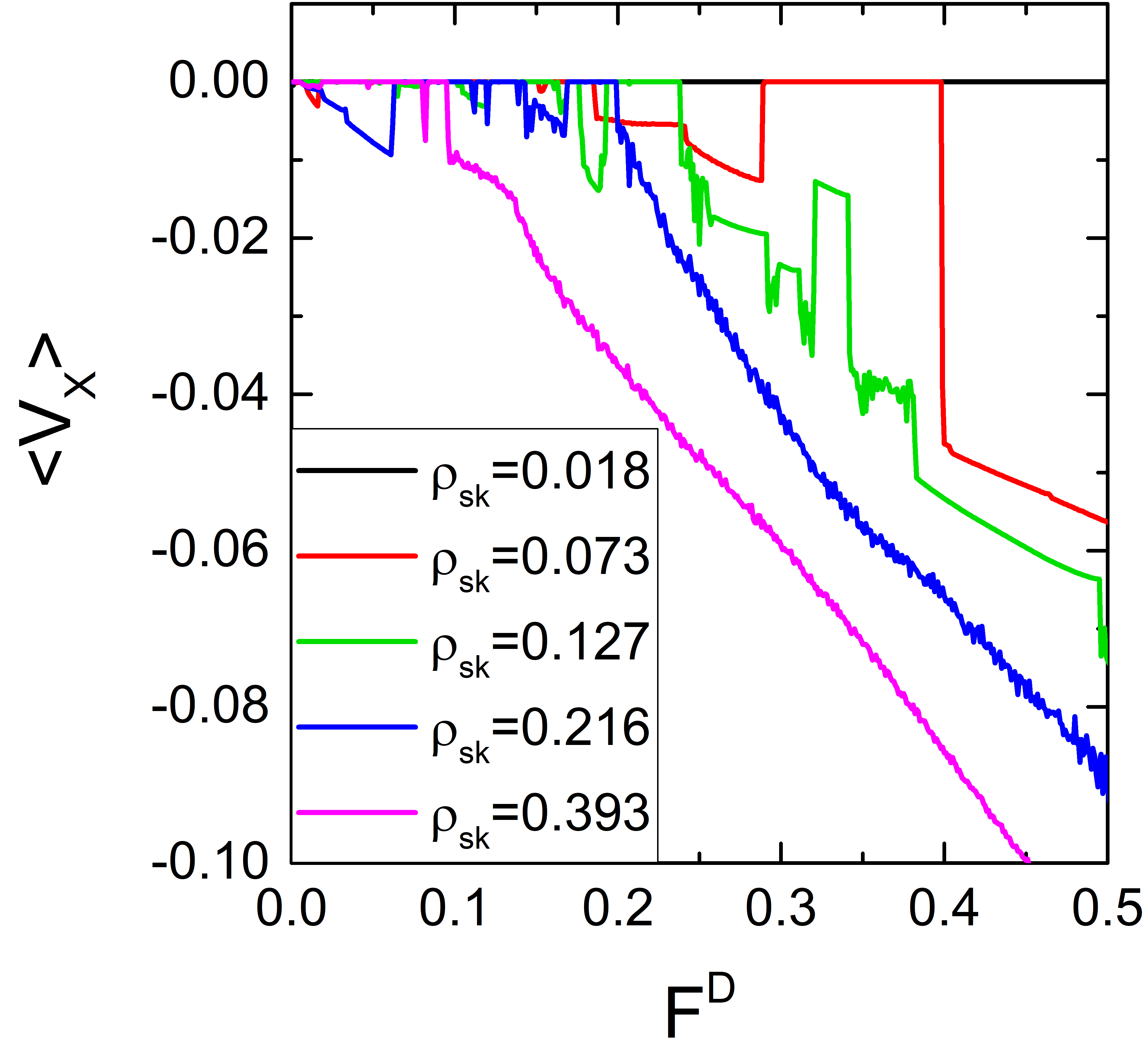}
  \end{center}
\caption{
A zoomed in view of
the $\langle V_x\rangle$ vs $F^D$ curves from Fig.~\ref{Fig4}(a) for the
system with $\alpha_m/\alpha_d=0.5$ and $-x$ direction driving
over the range $0<F^D<0.5$.
Reentrant pinning phases appear for all but the smallest value
of $\rho_{sk}$.
}
\label{Fig5}
\end{figure}

Under hard direction driving, 
$\langle V_x \rangle = 0$ for all values
of $F^D$
when $\rho_{sk} \leq 0.037$,
as shown in Fig.~\ref{Fig4}(b). 
A snapshot of the skyrmion positions
for $\rho_{sk}=0.018$ and $F^D=1.5$
appears in Fig.~\ref{Fig7}(a).
Skyrmions are trapped along the vertical walls of the funnel 
array and are too distant to interact with each other. 
When $\rho_{sk}>0.037$, the skyrmions begin to exhibit 
collective effects and finite
motion develops along the hard direction.
Interestingly, the magnitude of the average skyrmion velocity, $|\langle V_x \rangle|$, 
increases as the skyrmion density increases, as illustrated in Fig \ref{Fig4}(a). This is 
different from the behavior of overdamped particles 
\cite{reichhardt_clogging_2018,reichhardt_commensurability_2010},
where the average velocity 
decreases as the particle density increases.
In the skyrmion case, the Magnus force is 
present,
and it is well-known that gyroscopic effects can significantly influence the 
skyrmion organization \cite{brown_effect_2018}.
We believe the phenomenon here is analogous, 
and that as the density of skyrmions
increases, the gyroscopic term favors an easier flow of 
skyrmions.
We also find reentrant pinned phases (RPPs) for all systems
with $\rho_{sk}>0.037$,
as shown in Fig.~\ref{Fig5}. Note that the size and
number of RPPs vary with the skyrmion density.
For high skyrmion densities, such as 
$\rho_{sk}=0.393$, the RPPs are very narrow.
On the other hand, for low densities such as 
$\rho_{sk}=0.073$,
there is only one RPP and it extends over a fairly wide range of $F^D$. 
For intermediate skyrmion densities, such as
$\rho_{sk}=0.127$, the RPPs are numerous but
are less stable than those found at lower densities. 
    
\begin{figure}[ht]
\begin{center}
        \includegraphics[width=0.4\columnwidth]{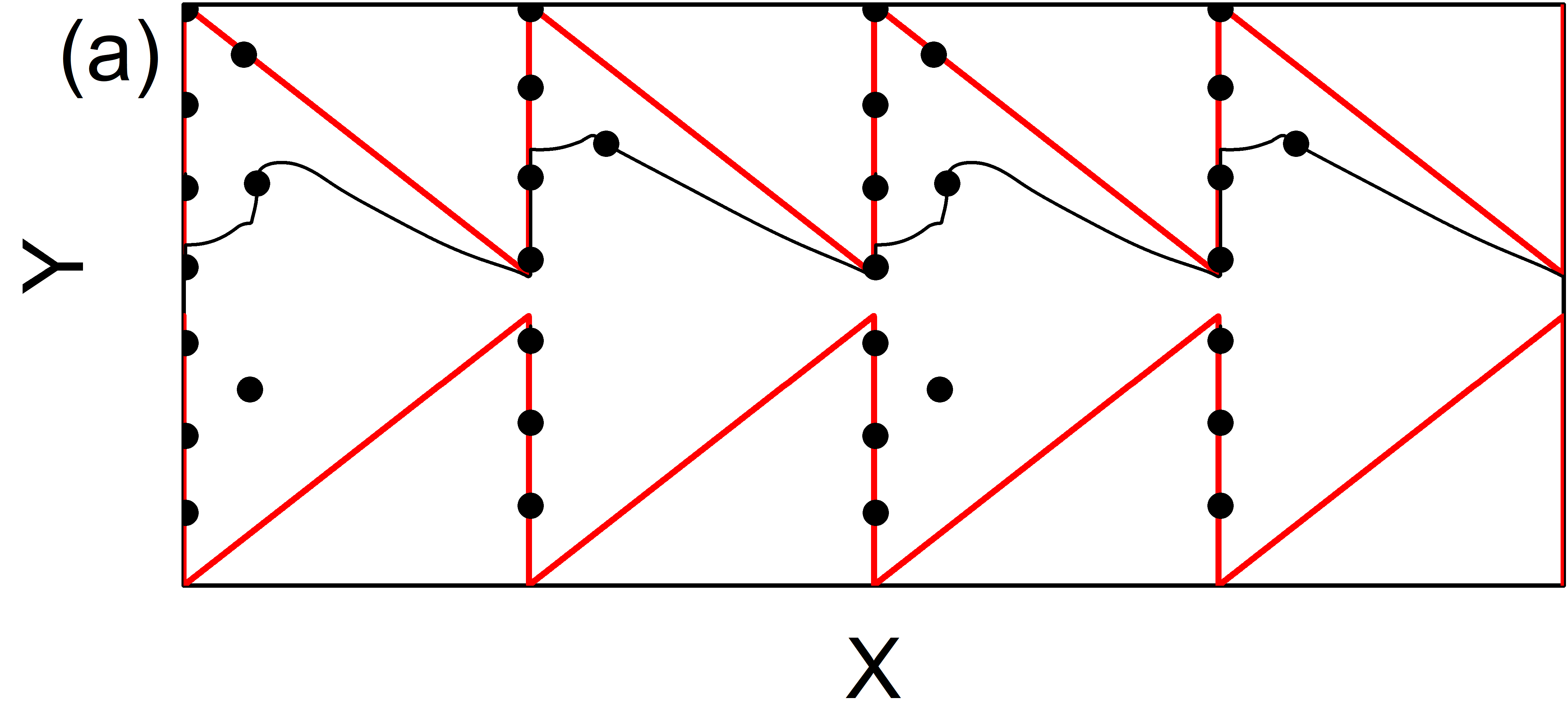}
        \includegraphics[width=0.4\columnwidth]{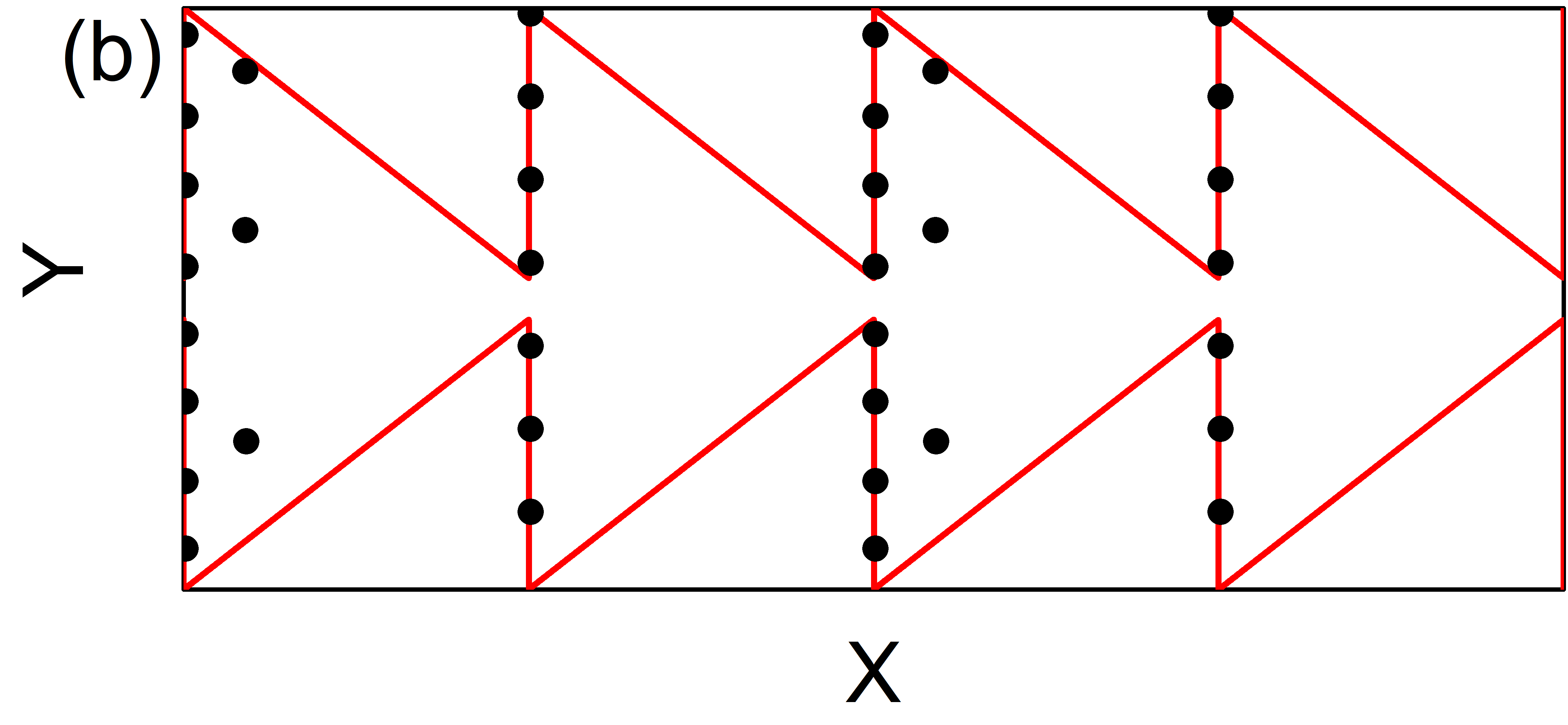}
        \includegraphics[width=0.4\columnwidth]{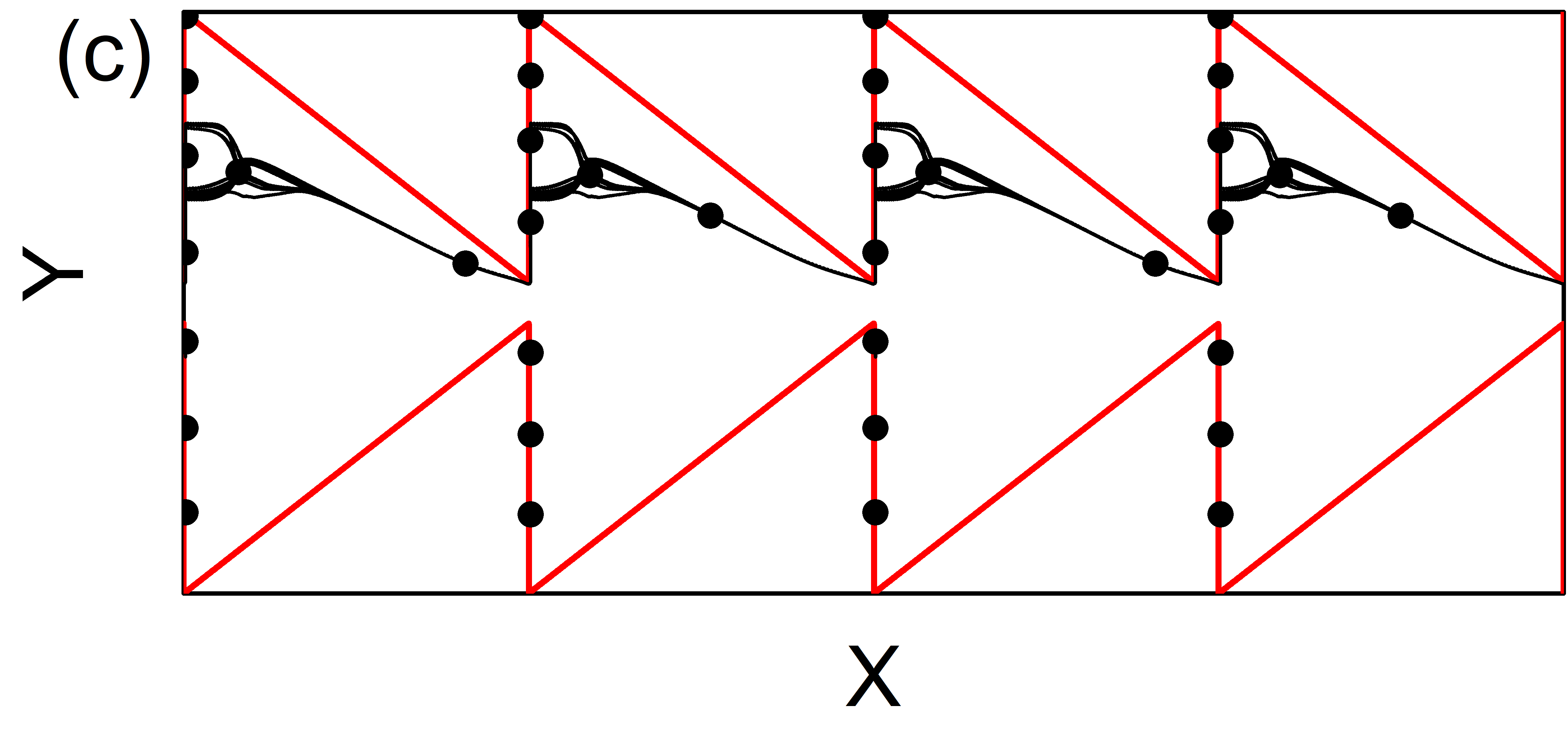}
\end{center}        
\caption{
Skyrmion trajectories for the system in Fig.~\ref{Fig4} with $-x$ driving
and $\alpha_m/\alpha_d=0.5$ at  
$\rho_{sk}=0.073$.
Red lines are the funnel walls,
black dots are skyrmions, and black lines are the skyrmion trajectories.
(a) For $F^D=0.26$, some of the skyrmions flow through the funnel
openings, resulting in finite values of $\langle V_x \rangle$.
(b) The reentrant pinned or clogged phase at $F^D=0.35$,
where skyrmions accumulate behind a blockage
caused when other skyrmions
become trapped on the vertical funnel wall,
decreasing the
effective width of the funnel opening
through their repulsive interactions.
(c) The sliding phase at $F^D=0.4$, where skyrmions are able to flow through the funnel opening again.
}
\label{Fig6}
\end{figure}

In Fig.~\ref{Fig6} we plot the skyrmion trajectories for
the system
from Fig.~\ref{Fig4}(a) and Fig.~\ref{Fig5}
with
$\alpha_m/\alpha_d = 0.5$ at
$\rho_{sk}=0.073$.
For $0.185<F^D<0.29$,
the skyrmions depin and start 
flowing in the $-x$ direction, as illustrated
in Fig.~\ref{Fig6}(a) for $F^D=0.26$.
In the interval 
$0.29<F^D<0.398$,
the skyrmion motion is interrupted by a \textit{clogging effect}, where 
skyrmions pile up in two of the funnel plaquettes
and block the flow, as illustrated in Fig.~\ref{Fig6}(b)
for $F^D=0.35$.
This leads to a zero velocity regime where the
skyrmions are trapped.
For $F^D>0.398$,
a second depinning transition occurs
and the skyrmions start to flow again, as 
shown in Fig.~\ref{Fig6}(c) for $F^D=0.4$. 
    
In Fig.~\ref{Fig7} we show some snapshots of the skyrmion positions
for the system from Fig. \ref{Fig4}.
As the skyrmion density 
increases,
skyrmions accumulate along the vertical walls of the
funnel array, as illustrated in Fig.~\ref{Fig7}(a,b).
Moreover, due to the
finite skyrmion Hall angle, skyrmions tend to accumulate in the upper part of 
the funnels, as shown in Fig.~\ref{Fig7}(c,d).
The hard direction flow of skyrmions
occurs
through two types of motion.
At low densities, a skyrmion arriving at the
upper portion of the vertical wall displaces the
skyrmions already stuck along the wall and pushes one of them into the
orifice, allowing it to flow into the next plaquette where the process
is repeated.
For higher densities,
the accumulation of skyrmions in the upper portion of each funnel
confines the skyrmion motion to a channel along the center line of the
funnel, and the wall displacement mechanism is disrupted.
Meanwhile, some skyrmions remain permanently trapped
in the lower part of the funnel.

\subsection{Skyrmion Diode Effect}
\begin{figure}[ht]
\begin{center}
    \includegraphics[width=0.35\columnwidth]{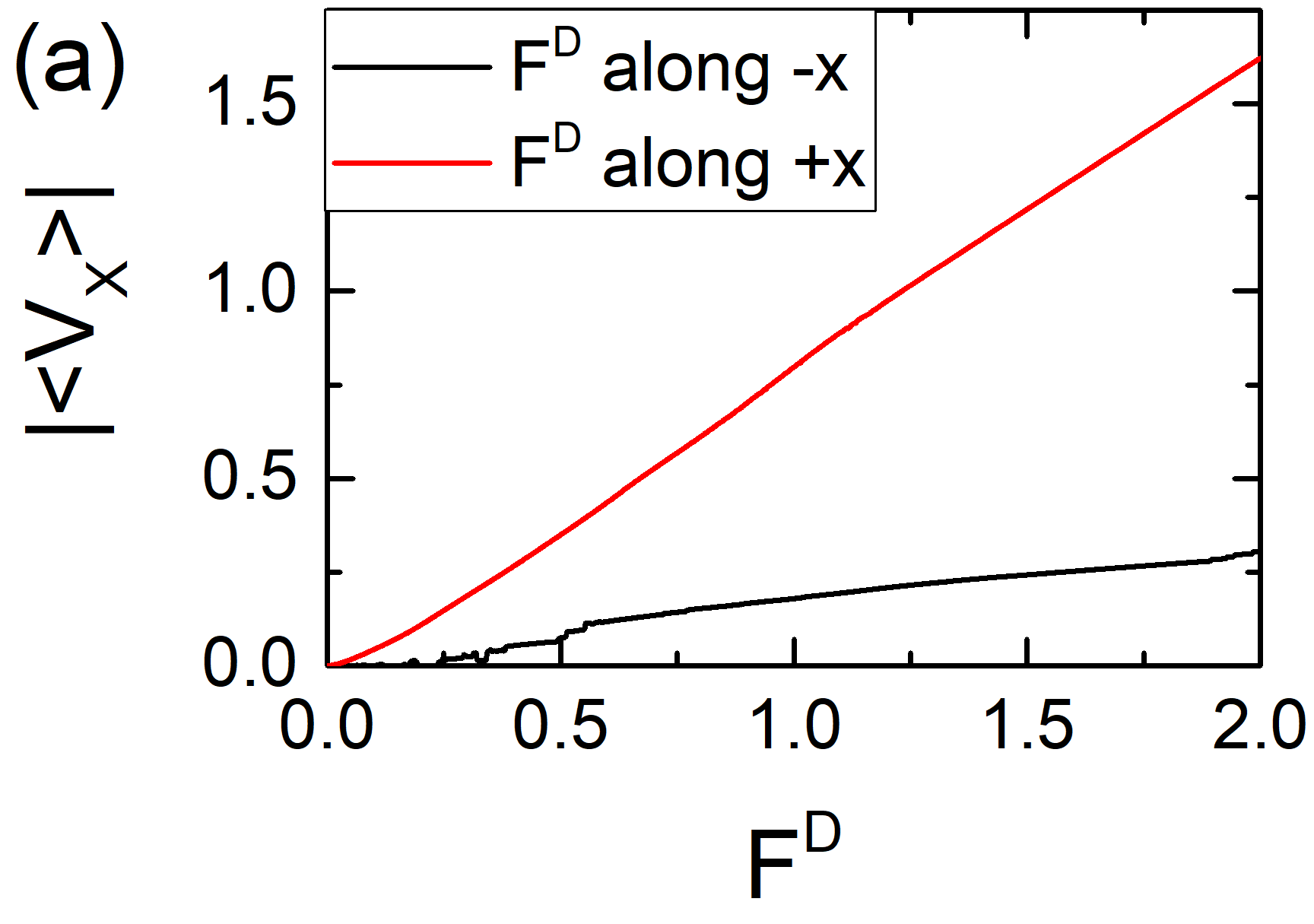}%
    \includegraphics[width=0.35\columnwidth]{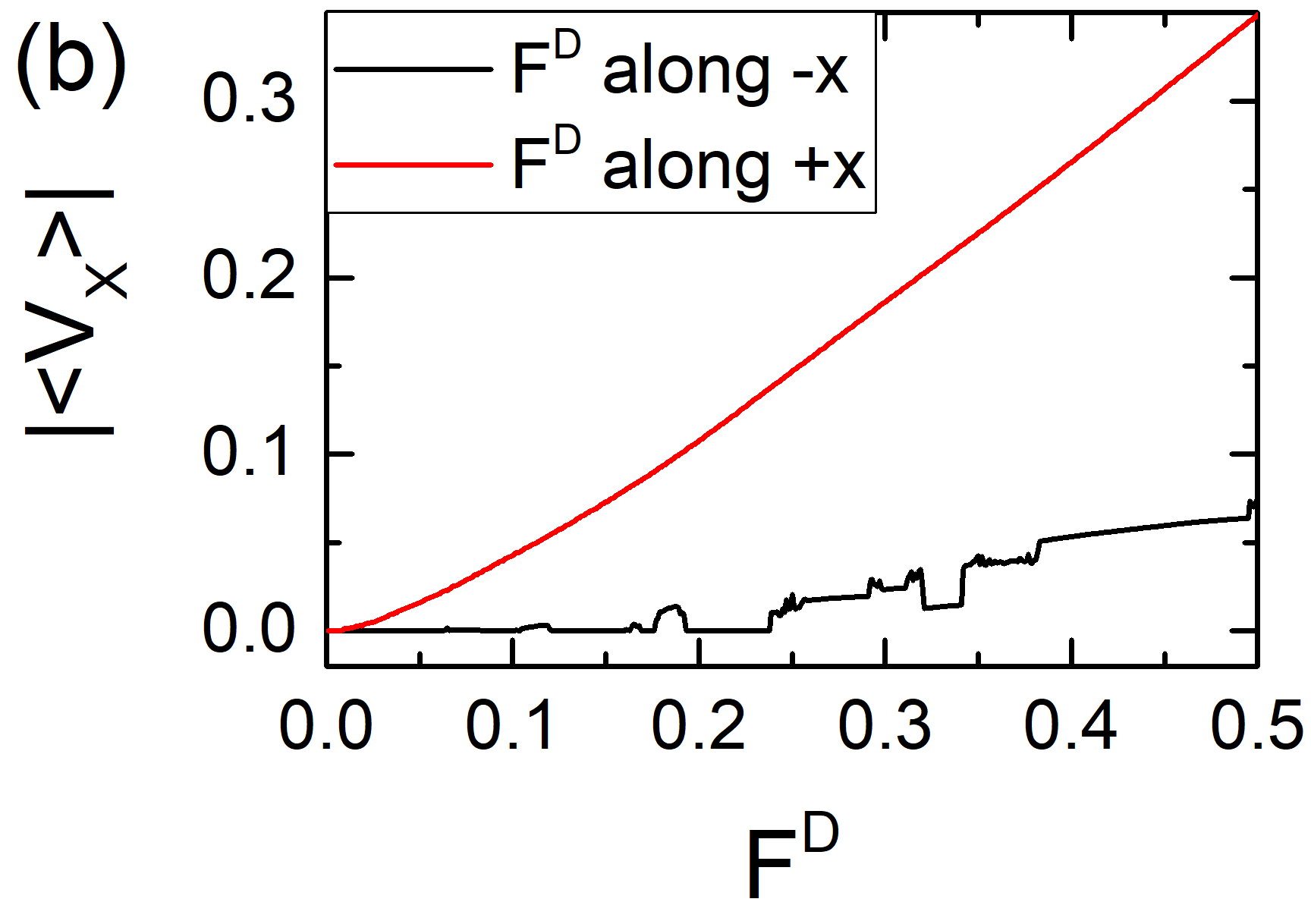}
\end{center}
\caption{
(a) $|V_{x}|$ vs $F_{D}$ for the system
from Figs.~\ref{Fig2} and \ref{Fig4} with $\alpha_m/\alpha_d=0.5$ and
$\rho_{sk}=0.127$ for driving in the $+x$ (red) and $-x$ (black)
directions.
(b) A zoomed in version of the same plot.
There is a skyrmion diode effect
in which the critical depinning force is finite only for
$-x$ driving, while the velocity response is much higher
for $+x$ driving.
}
\label{NewFig1}
\end{figure}

The difference in the velocity response
for driving in the easy and hard directions
can be viewed as an example of
a skyrmion diode.
In diodes, the threshold for 
transport is different
for one direction of drive compared to the other, and the diode effect
can be used to create a variety of logic devices
\cite{Katz05,Kitai11}.
A number of 
skyrmion diode
devices have been proposed using
voltage controlled potential barriers \cite{Zhao20a}, asymmetric
substrates \cite{Jung21},
and even the asymmetry
of the motion from the Hall effect \cite{Feng22}.
In our system, the diode effect arises because the
skyrmions are more susceptible to clogging for hard direction driving
than for easy direction driving.
To illustrate the diode effect more clearly,
in Fig.~\ref{NewFig1}(a) we plot 
$|V_{x}|$ versus $F_{D}$
for driving in the $+x$ and $-x$ directions for the system 
from Figs.~\ref{Fig2} and \ref{Fig4}
with $\alpha_{m}/\alpha_{d} = 0.5$
at $\rho_{sk} = 0.127$. 
There is a smooth increase in the velocity-force curve
and a depinning force of $F_{c} = 0$
for
$+x$ driving,
while for $-x$ driving,
the depinning force is
finite and has a value near $F_{c} = 0.25$, as 
shown in the zoomed in plot of Fig.~\ref{NewFig1}(b).
For driving in the
hard direction, a clogged regime appears
for $0.25 < F_{D} < 0.5$.
When $F_{D} > 0.5$, the velocity 
is always higher for $+x$ driving
than for $-x$ driving
since all of the skyrmions are in motion under easy direction driving,
but a portion of the skyrmions become trapped in the funnel corners
for hard direction driving.
Figure~\ref{NewFig1} indicates that the velocity-force
curves for $+x$ driving remain
nonlinear due to the collective effects of the skyrmions moving near
the funnel tip.
Within our particle-based model,
skyrmions that become trapped at the funnel 
edges for hard direction driving
remain trapped even if the driving force is increased;
however, for a real system or in a
continuum model, these skyrmions could eventually
become strongly distorted and depin for sufficiently large drives.

\section{Effect of varying $\alpha_m/\alpha_d$ for easy direction driving}

As shown in previous sections, the Magnus force plays a major role in
determining the dynamics of the skyrmions
confined to
the funnel array.
We next 
vary the ratio
$\alpha_m/\alpha_d$ to change the relative importance of the Magnus
term in a system with fixed $\rho_{sk}=0.127$.

\begin{figure}[ht]
\begin{center}
  \includegraphics[width=0.35\columnwidth]{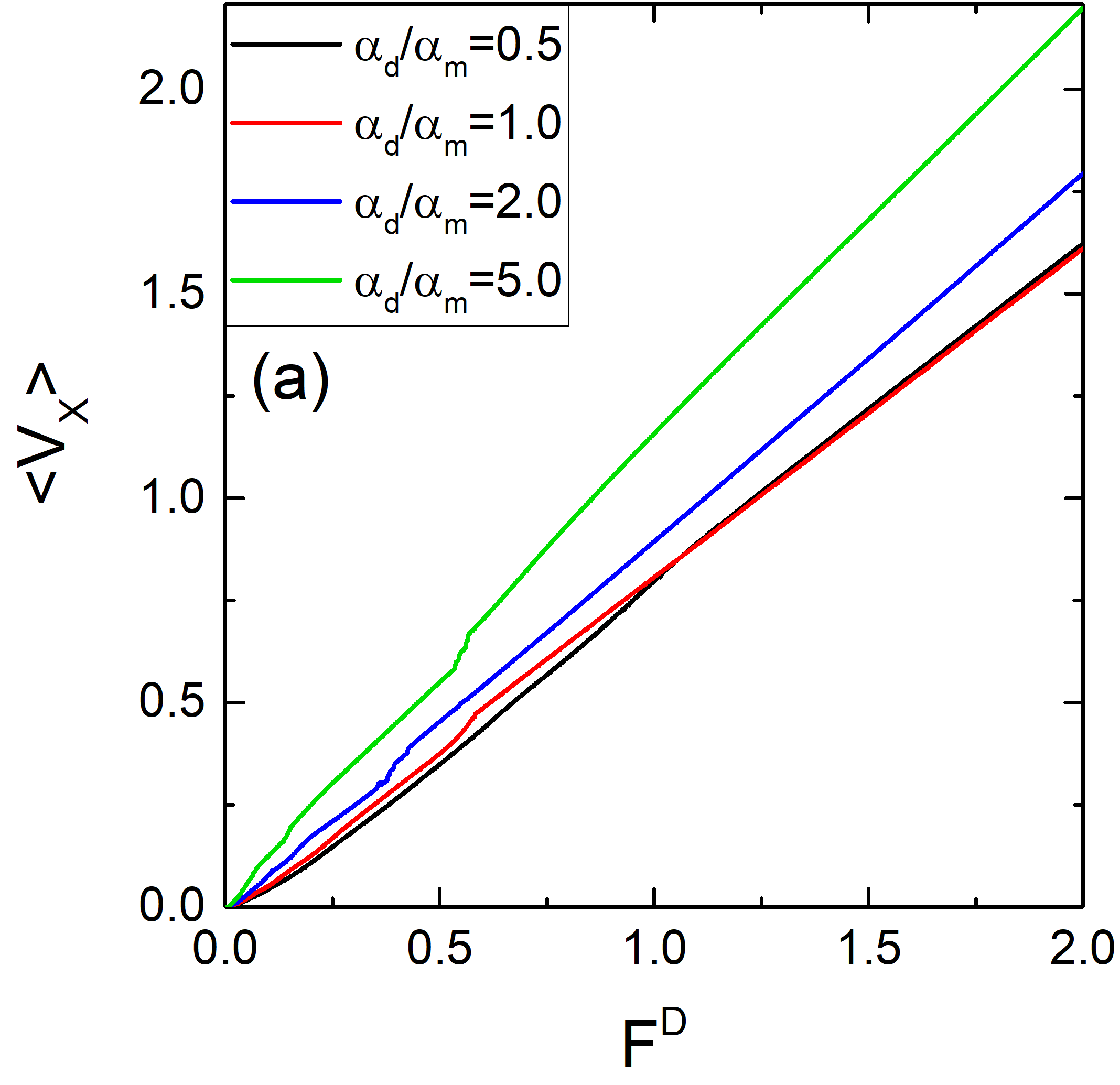}%
  \includegraphics[width=0.35\columnwidth]{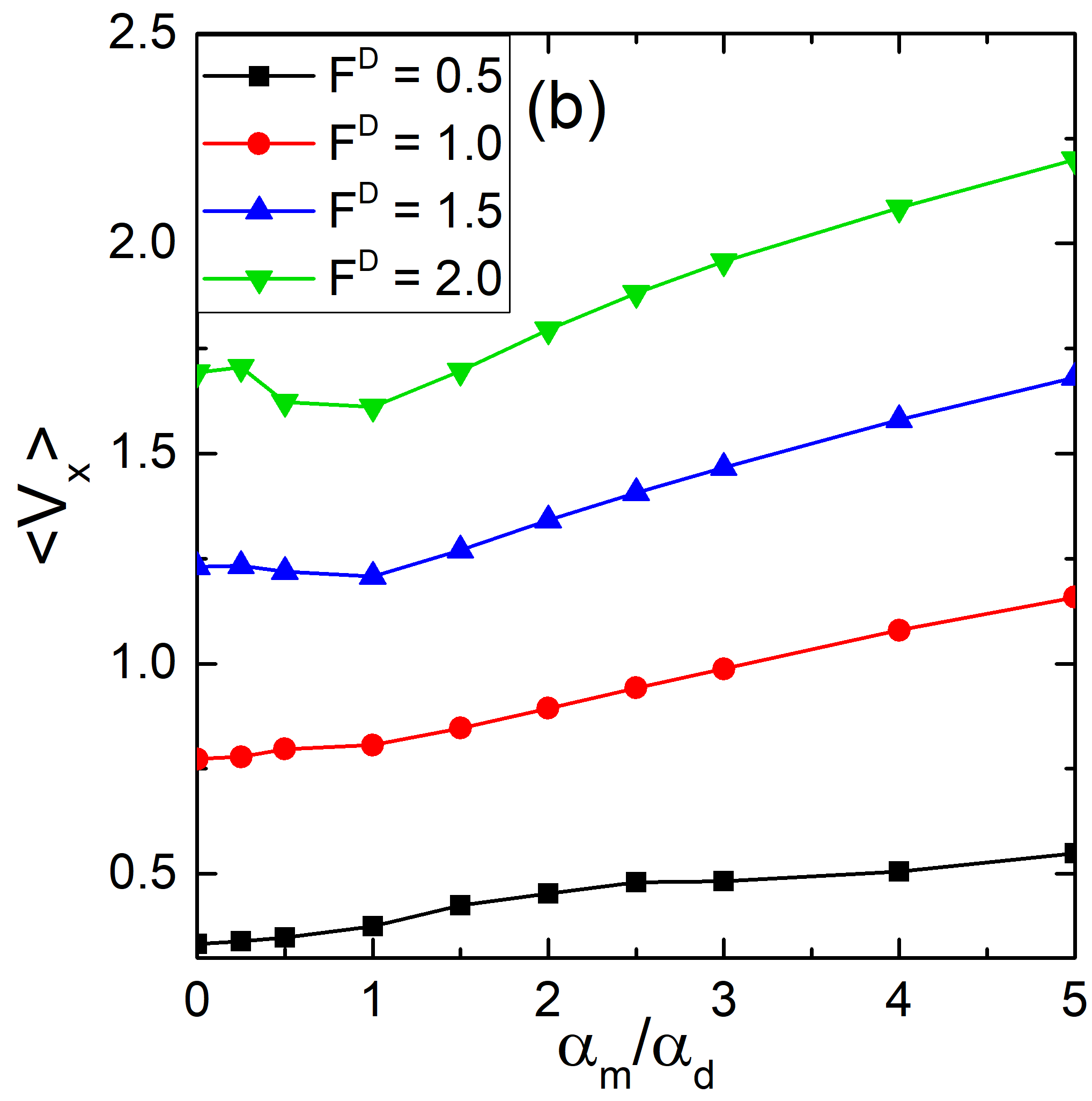}
\end{center}  
\caption{
Results for a sample with $\rho_{sk} = 0.127$ under $+x$ driving.
(a) $\langle V_x\rangle$ vs $F^D$ for 
$\alpha_m/\alpha_d=$0.5, 1.0, 2.0, and 5.0. 
(b) $\langle V_x \rangle$ vs $\alpha_m/\alpha_d$ for $F^D=$0.5, 1.0,
1.5, and 2.0.
}
\label{Fig8}
\end{figure}

In Fig.~\ref{Fig8}(a), the plots of $\langle V_x\rangle$ versus $F^D$ for
samples with $\rho_{sk}=0.127$ and varied $\alpha_m/\alpha_d$ under
$+x$ driving indicate
that the Magnus term modifies the average 
skyrmion velocity.
For stronger Magnus forces of $\alpha_m/\alpha_d >1.0$,
the $\langle V_x\rangle$ versus $\alpha_m/\alpha_d$
curves in Fig.~\ref{Fig8}(b) show that
$\langle V_x\rangle$
always increases as the Magnus force is increased. 
When $\alpha_m/\alpha_d \leq 1.0$, however, the behavior changes.
In the high drive regime, such as at $F^D = 2.0$,
$\langle V_x \rangle$ is nonlinear and non-monotonic
over the interval 
$0<\alpha_m/\alpha_d<1.0$.
In contrast,
for low and intermediate drive regimes, 
$\langle V_x \rangle$ increases monotonically with increasing Magnus force.

\begin{figure}[ht]
\begin{center}
  \includegraphics[width=0.6\columnwidth]{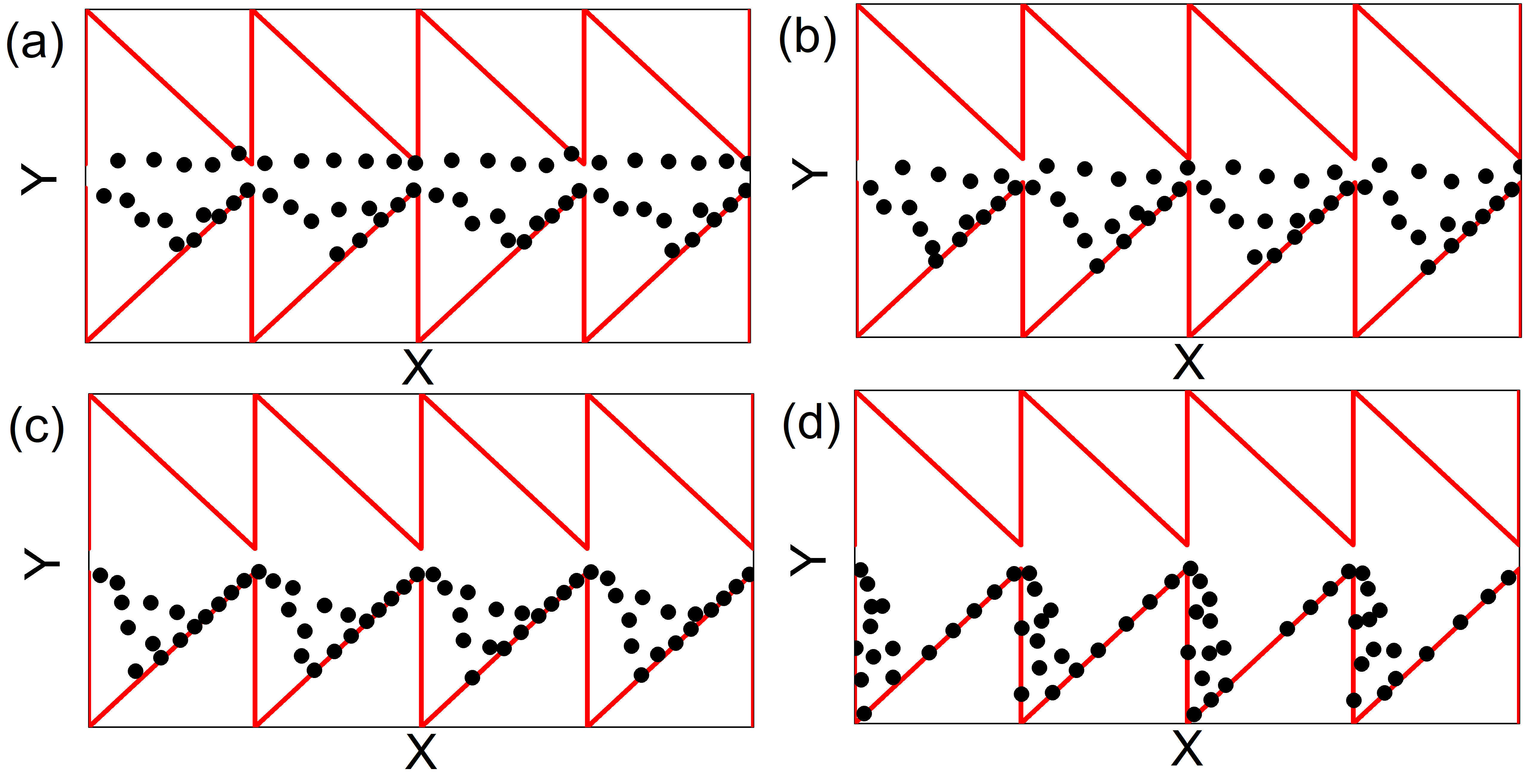}
\end{center}  
\caption{
Snapshots of the skyrmion positions for the system in
Fig.~\ref{Fig7} with $\rho_{sk}=0.127$ at $F^D=1.5$.
Red lines are the funnel walls
and black dots are skyrmions.
The Magnus force ratio is
$\alpha_m/\alpha_d=$ (a) 0.25, (b) 0.5, (c) 1.0 and (d) 5.0.
}
\label{Fig9}
\end{figure}

In Fig.~\ref{Fig9} we show some representative snapshots of
the skyrmion positions for the system
from Fig.~\ref{Fig8} at $F^D = 1.5$.
For 
$\alpha_m/\alpha_d = 0.25$
in Fig.~\ref{Fig9}(a),
where the Magnus force is low,
the
intrinsic skyrmion Hall angle
is $\theta_{sk}^{\rm int}=-14.04^{\circ}$ and the skyrmions
move along two channels.
In the lower channel, the skyrmions follow the direction of
the intrinsic skyrmion Hall 
angle until they strike
the diagonal funnel wall, which guides them to
the funnel opening. In
the upper channel, the skyrmions flow nearly parallel to the drive
because
the intrinsic skyrmion Hall angle is
canceled by the repulsive interactions
with skyrmions in the lower channel.
As both channels
approach the funnel opening,
the skyrmions compete to move into the next funnel plaquette.
In Fig.~\ref{Fig9}(b)
for $\alpha_m/\alpha_d = 0.5$, the intrinsic Hall angle is higher 
($\theta_{sk}^{\rm int}=-26.56^{\circ}$)
and therefore the skyrmions in the lower channel strike the edge
of the diagonal funnel wall further to the left.
When $\alpha_m/\alpha_d = 1.0$, as shown in Fig.~\ref{Fig9}(c),
$\theta_{sk}^{\rm int}=-45^{\circ}$ and the skyrmion channels
begin to merge before reaching the funnel opening,
producing a denser single channel.
In Fig.~\ref{Fig9}(d), where
$\alpha_m/\alpha_d = 5.0$ and $\theta_{sk}^{\rm int} =-78.7^{\circ}$,
there are no longer clearly separated channels
and the skyrmions flow in a single chaotic band of 
motion.
The presence of two channels of flow for lower Magnus forces
facilitates clogging of the skyrmions since the channels
compete as they approach the funnel
opening, while the single channel or band of flow at higher Magnus
forces has no competition and
can pass more easily from one funnel plaquette to the next,
increasing the average skyrmion velocity.

\begin{figure}[ht]
\begin{center}
  \includegraphics[width=0.4\columnwidth]{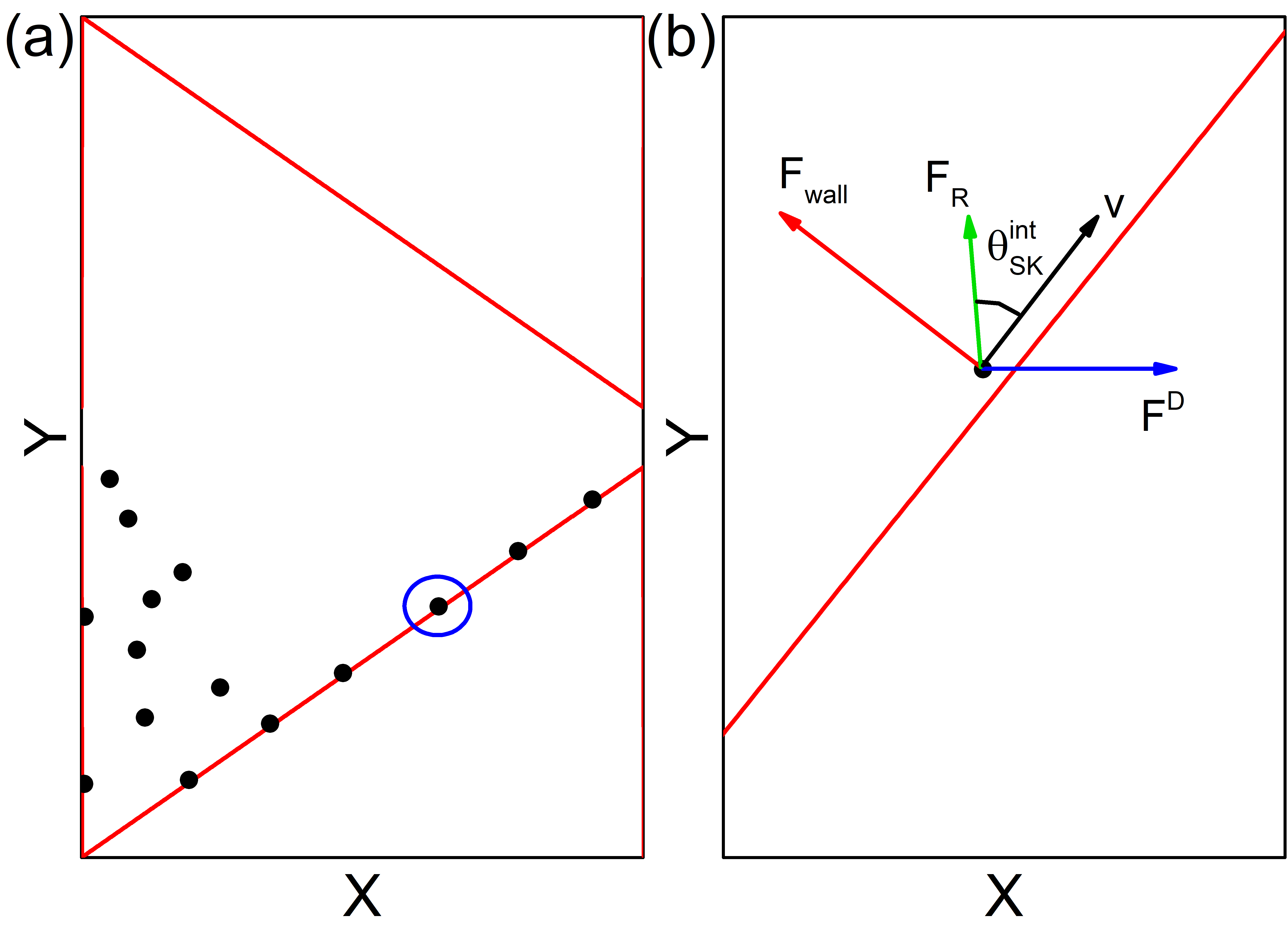}
\end{center}  
\caption{
(a) A zoomed in image of a single funnel from Fig.~\ref{Fig9}(d).
Red lines are the funnel walls
and black dots are skyrmions.
We focus on the circled skyrmion.
(b) Force diagram indicating the force components that contribute
to the velocity boost experienced by
a skyrmion in the direction parallel to the
funnel wall.
$\mathbf{F}_D$ (blue) is the driving force in the $+x$ direction.
$\mathbf{F}_{\rm wall}$ (red) is the repulsive force from the
funnel wall, which acts perpendicular to the wall.
$\mathbf{F}_R$ (green) is the sum of $\mathbf{F}_D$ and $\mathbf{F}_R$.
The Magnus force rotates $\mathbf{F}_R$ clockwise
by an amount $\theta_{sk}^{\rm int}=\arctan(\alpha_m/\alpha_d)$.
In the illustrated case, this angle exactly matches the angle of the
funnel wall, and the entire force $\mathbf{F}_R$ acts in the direction
of the skyrmion velocity $\mathbf{v}$ (black) along the funnel wall,
giving a maximal velocity boost effect.
In more general cases, the skyrmion Hall angle and funnel wall angles
are different, and only a portion of $\mathbf{F}_R$ contributes
to the boost effect.
}
\label{Fig10}
\end{figure}

In Fig.~\ref{Fig9}(d) we find that the skyrmions along the 
diagonal funnel wall are more widely spaced
than the other skyrmions. This is caused by
the velocity boost that the skyrmions receive
when they are flowing against
the wall.
Figure~\ref{Fig10} describes the
origin of the velocity boost effect schematically.
In Fig.~\ref{Fig10}(a) we highlight a single funnel
where the high speed flow along the funnel wall occurs, with one
skyrmion circled.
The forces on this skyrmion
are illustrated in Fig.~\ref{Fig10}(b).
The driving force $\mathbf{F}_D$ is in the $+x$ direction,
and the repulsive force from the wall $F_{\rm wall}$ acts
perpendicular to the wall.
The vector sum of these two forces is
$\mathbf{F}_R$. According to equation (\ref{vels}), the Magnus force rotates
$\mathbf{F}_R$ clockwise by an amount $\theta_{sk}^{\rm int}$,
toward the direction of $\mathbf{v}$. Depending on the particular value of
$\alpha_m/\alpha_d$, this will bring a greater or lesser portion of
$\mathbf{F}_R$ into alignment with the skyrmion velocity $\mathbf{v}$ parallel
to the funnel wall, producing a velocity boost. In the illustrated case,
$\theta_{sk}^{\rm int}$ matches the angle of the funnel wall
exactly and the
velocity boost effect is maximized.

    \section{Effect of varying $\alpha_m/\alpha_d$ for hard direction driving}

\begin{figure}[ht]
 \begin{center}
    \includegraphics[width=0.35\columnwidth]{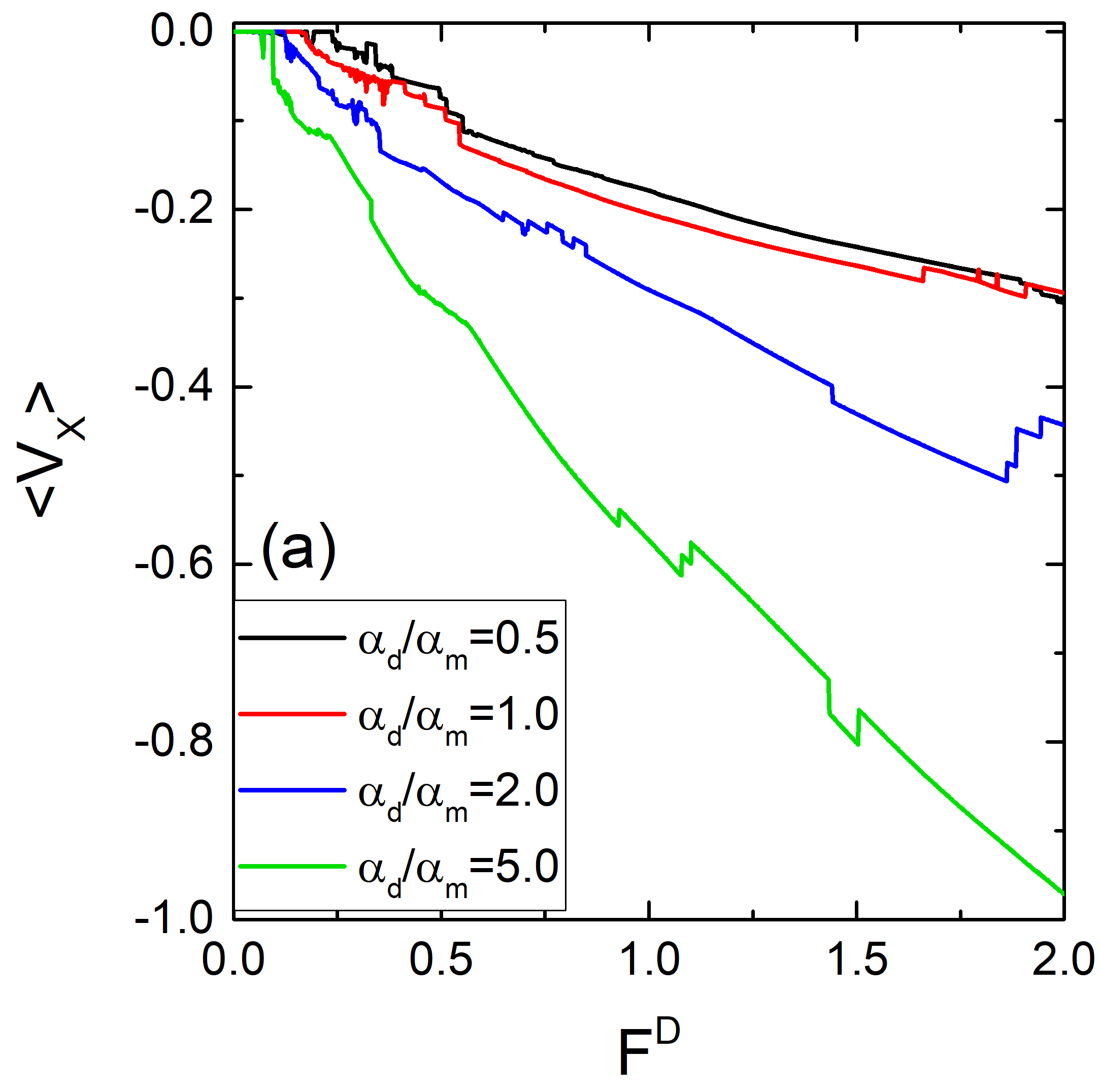}%
    \includegraphics[width=0.35\columnwidth]{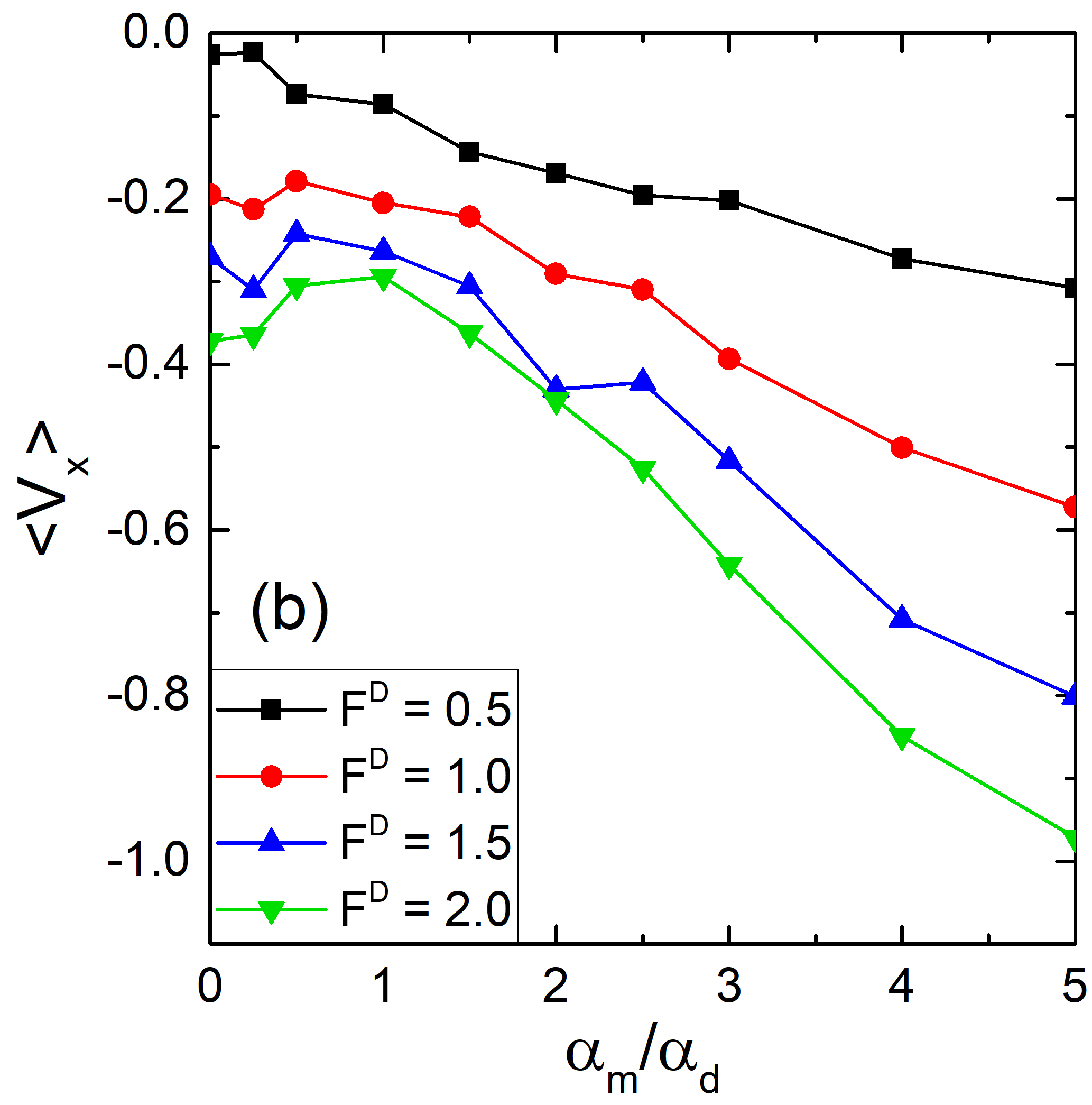}
 \end{center}
\caption{
Results for a sample with $\rho_{sk} = 0.127$ under $-x$ driving.
(a) $\langle V_x\rangle$ vs $F^D$ for $\alpha_m/\alpha_d$=0.5,
1.0, 2.0, and 5.0.
(b) $\langle V_x \rangle$ vs $\alpha_m/\alpha_d$ for $F^D$ = 0.5, 1.0, 1.5,
and 2.0.
}
\label{Fig11}
\end{figure}

We now apply the drive along the hard or $-x$ direction while
varying $\alpha_m/\alpha_d$ for samples with $\rho_{sk}=0.127$.    
In Fig.~\ref{Fig11}(a), we plot $\langle V_x\rangle$ versus $F^D$ for
different values of $\alpha_m/\alpha_d$.
There is an overall increase in
the magnitude of the average skyrmion velocity as the
Magnus term becomes larger, which is more clearly illustrated in the
plot of $\langle V_x\rangle$ versus $\alpha_m/\alpha_d$ for selected
values of $F^D$ in Fig.~\ref{Fig11}(b).
Note that reentrant pinning phases (RPPs) are also affected by the Magnus 
intensity.
The RPPs are more prominent for low values of 
$\alpha_m/\alpha_d$
but the clogging effect
is diminished as the Magnus term becomes stronger.
    
Figure~\ref{Fig11}(a) shows that 
the depinning threshold changes as $\alpha_m/\alpha_d$ is varied. 
For higher values of $\alpha_m/\alpha_d$, 
the depinning threshold shifts to lower values of $F^D$.
In Fig.~\ref{Fig11}(b), $\langle V_x \rangle$ has a
nonlinear and non-monotonic
dependence on $\alpha_m/\alpha_d$ for intermediate 
and high drives. 
This is similar to the results found for $+x$ driving in
Fig.~\ref{Fig8}.

\begin{figure}[ht]
 \begin{center}
   \includegraphics[width=0.6\columnwidth]{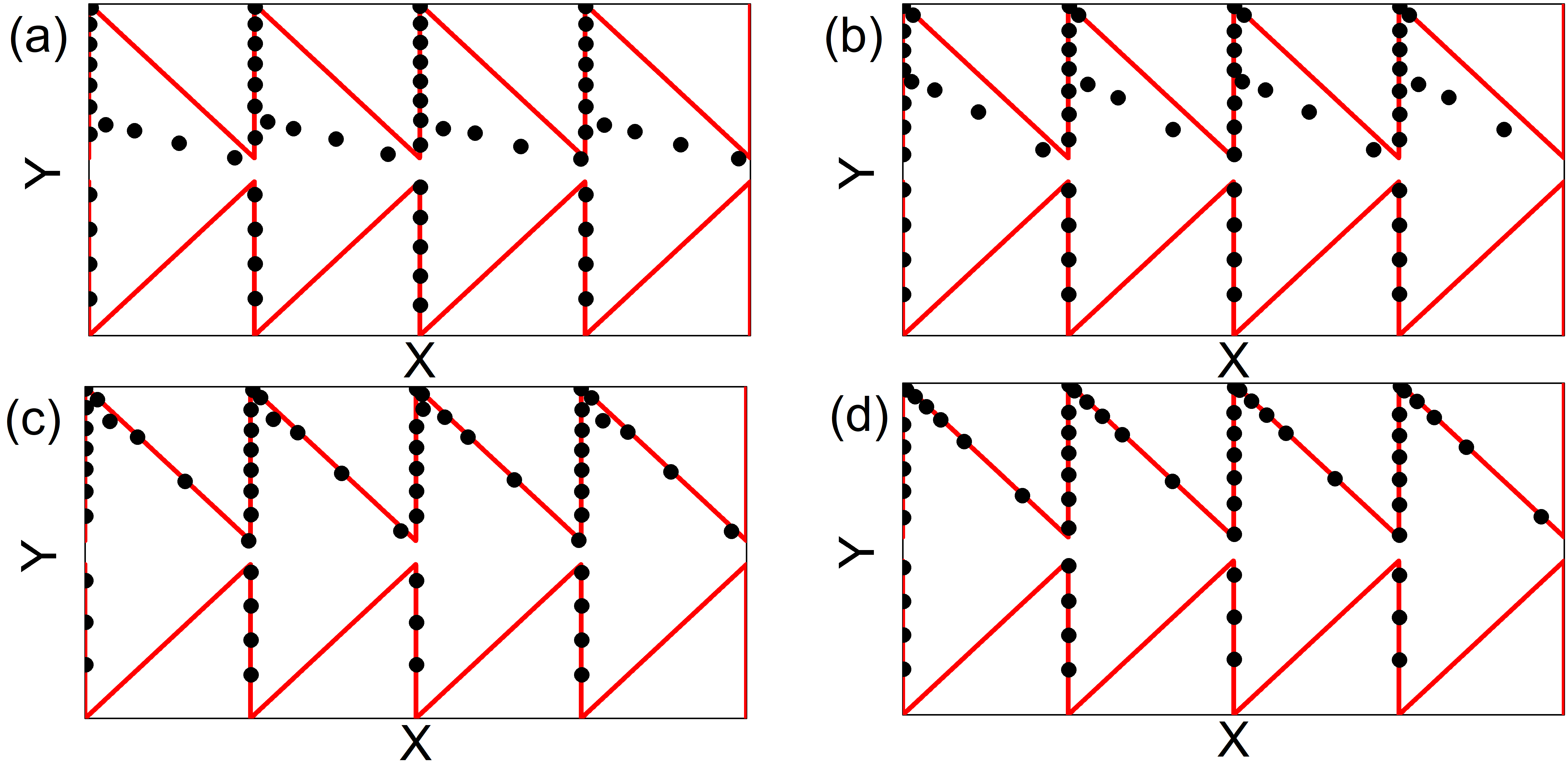}
\end{center}   
\caption{
Snapshots of the skyrmion positions for the system in Fig.~\ref{Fig11}
with $\rho_{sk}=0.127$ at $F^D=1.5$.
Red lines are the funnel walls and black dots are skyrmions.
The Magnus force ratio is $\alpha_m/\alpha_d=$ (a) 0.25, (b) 0.5, (c) 1.0 and (d) 5.0.
}
\label{Fig13}
\end{figure}

In Fig.~\ref{Fig13}(a) we illustrate snapshots of the
skyrmion positions for the system in Fig.~\ref{Fig11} at $F^D=1.5$.
For low values of $\alpha_m/\alpha_d$, the intrinsic skyrmion Hall 
angle is reduced and as a consequence more
skyrmions remain trapped in the funnel plaquettes.
For example, in Fig.~\ref{Fig13}(a) at $\alpha_m/\alpha_d = 0.25$,
the channel of flowing
skyrmions strikes the vertical funnel wall just above the point where
the bottom skyrmion is trapped by the wall, releasing that skyrmion and
allowing it to pass into the next funnel plaquette; however, all of the
other skyrmions higher up on the funnel wall remain trapped
permanently.
Since relatively few skyrmions are able to move,
the
average skyrmion velocity
is reduced.
For $\alpha_m/\alpha_d = 0.5$, shown in Fig. \ref{Fig13}(b),
the skyrmion channel strikes the vertical funnel wall approximately at its
center,
pushing a larger number of skyrmions
toward the funnel opening and reducing the
number of skyrmions that
are permanently pinned.
In Figs.~\ref{Fig13}(c,d) for $\alpha_m/\alpha_d = 1.0$ and $5.0$, 
respectively, the moving skyrmion channel
pushes almost all of the skyrmions along the vertical funnel wall
toward the funnel opening,
leading to a much higher average skyrmion velocity.
In addition, according to
the mechanism explained in Fig.~\ref{Fig10},
the skyrmion velocity for sliding along the diagonal funnel walls
can be boosted considerably.
The channel motion
of skyrmions boosted by the Magnus 
effect also
reduces the depinning threshold,
as shown in Fig.~\ref{Fig11}(a).

\section{Clogging Dynamics}
In systems that exhibit clogging,
such as flow though apertures or bottlenecks, the flow is
typically probabilistic in nature,
and the particles may flow for some period of time before reaching
a clogged configuration \cite{redner_clogging_2000,zuriguel_clogging_2014,to_jamming_2001,helbing_simulating_2000,nguyen_clogging_2017}.
In contrast,
systems that exhibit jamming have a unique density above which
the flow ceases \cite{Stoop18,Peter18}. 
In our system, the initial
skyrmion configurations are obtained using simulated annealing.
If somewhat different initial conditions
are used, the clogging behavior
found for $-x$ driving remains robust but the
specific drives at which
velocity dips or reentrant pinning
occur can shift to slightly different locations
for different
realizations.
For driving in the easy direction, however, the 
velocity curves remain unchanged 
for different initial configurations. 

\begin{figure}[ht]
  \begin{center}
    \includegraphics[width=0.4\columnwidth]{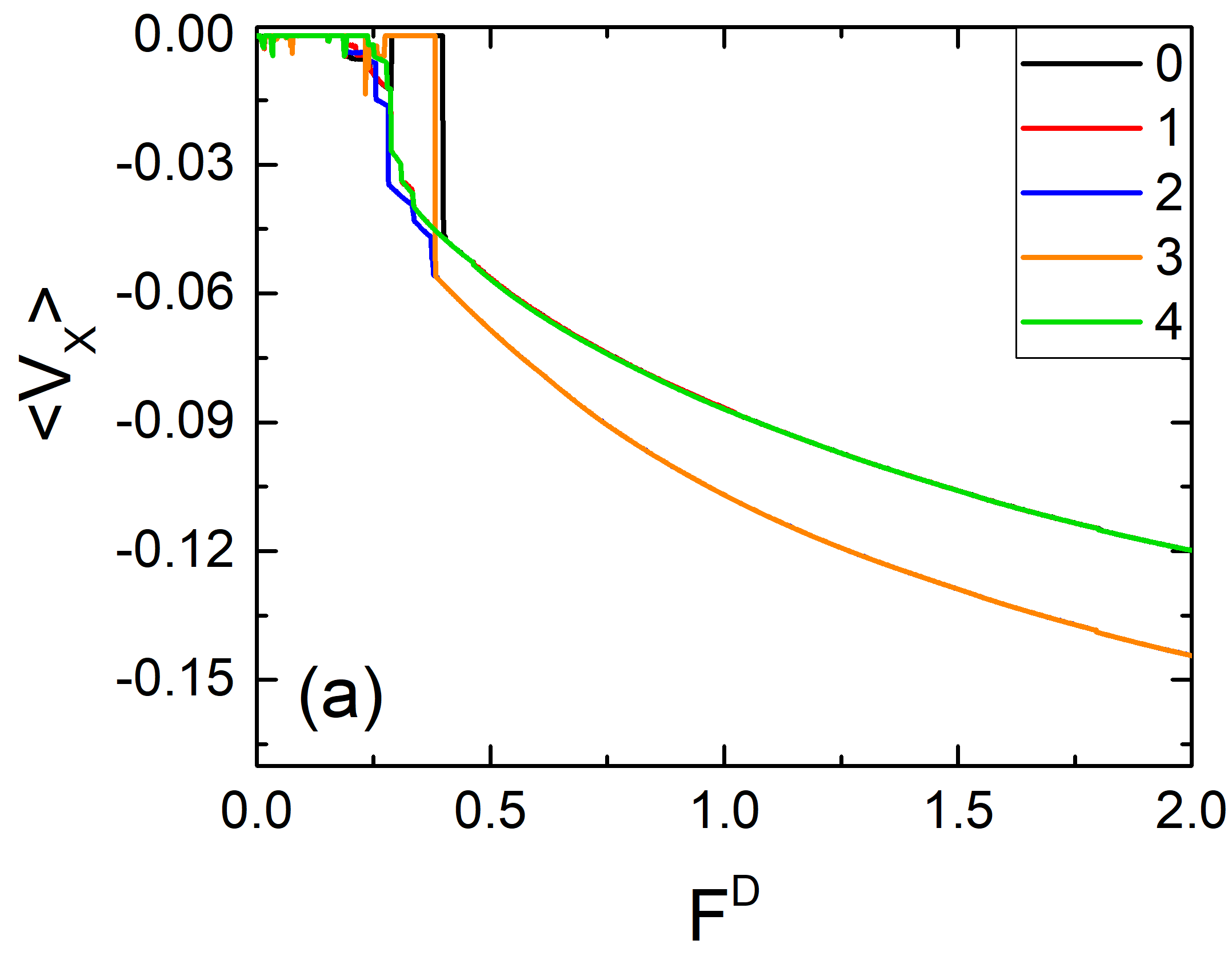}%
    \includegraphics[width=0.36\columnwidth]{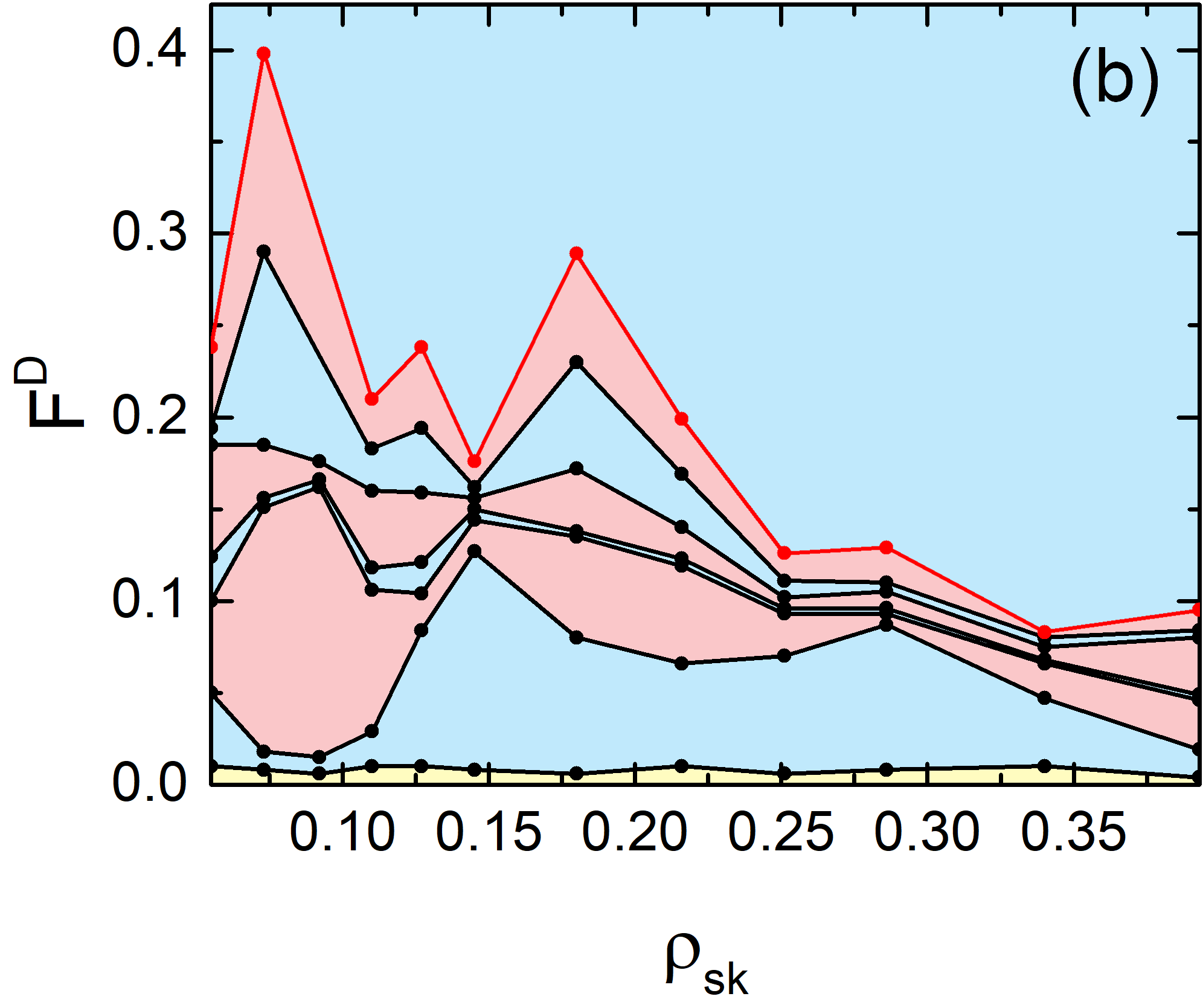}
    \end{center}
\caption{
(a) Results for a sample with $\rho_{sk}=0.127$ under $-x$ driving at
$\alpha_{m}/\alpha_{d} = 0.5$ for five different realizations of the
initial conditions.
The clogging events occur at slightly different drives,
while for $F^D > 0.45$, the system exhibits continuous flow that is
not sensitive to the initial conditions.
(b) Dynamic phase diagram as a function of $F^D$ vs $\rho_{sk}$
showing pinned (yellow), clogged (red), and flowing (blue) states for the
system in panel (a).
Below the red solid line, the system shows clogging behavior,
while above the red solid line, the flow is continuous.
}
\label{Fig15}
\end{figure}

In Fig.~\ref{Fig15}(a) we plot
$\langle V_x\rangle$ versus $F^D$
for driving in the hard direction in a system
with $\rho_{sk} = 0.073$ and $\alpha_{m}/\alpha_{d} = 0.5$
for five different realizations of the
initial conditions.
For $F^D > 0.45$, the curves are smooth and are insensitive to the
initial conditions,
while for $F^D < 0.45$, there are some differences in the
location of jumps and drops in the velocity for
different realizations; 
however, all of the realizations have jumps
for $F^{D} < 0.45$,
indicating that there
is a well defined clogging regime.
In Fig.~\ref{Fig15}(b), we plot a dynamic phase diagram
as a function of $F^D$ versus $\rho_{sk}$
showing how the clogging regime evolves.
As the skyrmion density increases,
the windows of drive over which clogging  can
occur are reduced.
In general, we find that there is a region
in which some flow occurs before a clogged state is established.
In clogging systems, the
probability to clog also depends on the rate of flow through
the aperture, such that
for lower flow rates the particles can generally flow,
while at higher flow rates  clogging becomes more likely.
In the skyrmion system, at high $F^{D}$ the drive can break up the clogs. 
The boundaries in Fig.~\ref{Fig15}(b) are obtained using individual
realizations at each point.
For multiple realizations at each point,
the system would be better described
as showing clogging behavior below the red solid
line and flowing behavior above the red solid line.

\section{Discussion}
In our system we consider a particle-based model
in which the number of particles is fixed 
and cannot change.
For real skyrmion systems,
it is possible for skyrmions
to be created or annihilated due to interactions with the current
and with barriers.
Additionally, skyrmions
could change shape when interacting with structures or with the drive.
An interesting future direction would be to consider
continuum-based models for skyrmions and the skyrmion-barrier 
interactions \cite{iwasaki_universal_2013,Lin13,DelValle22}
where the shape of the skyrmion
could change as it moves though the funnel constrictions.  
In the clogging phase,
the skyrmions could also become compressed or change
shape, and it would be interesting to explore whether
such extra degrees of freedom would
enhance or decrease the clogging effects.
The effect of temperature could also be important, since
in some cases
skyrmions are known to exhibit Brownian like motion
\cite{Zhao20}.
In this case,
thermal effects on the transport curves for driving in the 
easy direction would be small;
however, in the clogging regime, thermal effects could
have a strong impact by
causing unclogging events,
or allowing
the velocity at a fixed drive to undergo thermally induced jumps
from clogged to unclogged states.
We have focused on ferromagnetic skyrmions; however, different effects could
arise for driving other kinds of skyrmion textures
such as antiferromagnetic skyrmions, merons, 
or elliptical skyrmions \cite{Gobel21a}. 
There are various methods that
could be used to create funnel structures experimentally,
such as etched samples with thickness modulations, the addition of
adatoms in 
certain patterns on the surface \cite{Arjana20},
or ion irradiation \cite{Juge21}.  

\section{Summary}

Using a particle based model we simulated the collective effects of skyrmions
interacting with a linear 
array of funnel potentials
at zero temperature
for external dc driving applied along the easy 
or hard direction.
For easy direction driving,
we find that at low skyrmion densities
the dynamics is very similar to
that observed in the single skyrmion limit,
with the average velocity increasing 
linearly with increasing drive.
As the skyrmion density increases,
collective effects become relevant,
and for high skyrmion densities
the skyrmions may jam near the funnel opening, obstructing the flow.
For hard direction driving,
skyrmion motion
occurs only above a critical skyrmion density 
$\rho_{sk}^{\rm crit}=0.037$,
where the interskyrmion distance becomes small enough for
collective effects to emerge.
For densities above the critical value,
the skyrmions flow
in a window of driving forces
before entering
a reentrant pinning phase (RPP)
where the skyrmion motion is 
interrupted
due to a pile up of skyrmions inside
funnel plaquettes, leading to the formation of a clogged state.
Skyrmion motion in the clogged state drops to zero until the drive
is increased above
a second depinning threshold,
where the blockage is destroyed and the skyrmions can flow
through the funnels again.
We also observe
a skyrmion diode effect in which
there is a finite critical depinning threshold for
hard direction driving
but a zero critical threshold for easy direction driving.
The velocity for hard direction
driving is always lower than
that found for easy direction driving.

In addition to studying the effect of changing the
skyrmion density,
we also analyze the effects of modifying the
relative importance of the Magnus term by varying the
ratio $\alpha_m/\alpha_d$.
For both directions of applied drive,
the skyrmion velocity has a nonlinear
and non-monotonic dependence on $\alpha_m/\alpha_d$.
Moreover, the Magnus term 
also determines the spatial
distribution
of the skyrmions
inside the funnels, since
the skyrmion Hall angle causes skyrmions to be displaced toward
the lower half of the funnel for easy direction driving and toward the
upper half of the funnel for hard direction driving.
Increasing the strength of the Magnus term
changes the RPPs found for hard direction driving by making it more
difficult for a clogged state of skyrmions to form.
    
%    \acknowledgments
\ack
    This work was supported by the US Department of Energy through the Los Alamos National Laboratory. Los
    Alamos National Laboratory is operated by Triad National Security, LLC, for the National Nuclear Security
    Administration of the U. S. Department of Energy (Contract No. 892333218NCA000001). 
    \\
    J.C.B.S and N.P.V acknowledges funding from Fundação de Amparo à Pesquisa do Estado
    de São Paulo - FAPESP (Grants 2021/04941-0 and 2017/20976-3 respectively).

\section*{References}
\bibliographystyle{iopart-num}   
\bibliography{refs}
\end{document}